\documentclass[pageno]{jpaper}

\usepackage[normalem]{ulem}
\usepackage[utf8]{inputenx}
\usepackage[T1]{fontenc}
\usepackage{tikz}
\usepackage{amsmath}
\usepackage{amssymb}
\usepackage[EULERGREEK]{sansmath}
\usepackage{inconsolata}
\usepackage{caption}
\usepackage{filecontents}
\usepackage{mathtools}
\usepackage[toc,page]{appendix}

\usepackage[capitalise,noabbrev]{cleveref}
\usepackage{soul}
\usepackage{tabularray}
\usepackage{arydshln}
\usepackage[linesnumbered,ruled]{algorithm2e}
\Crefname{algocf}{Algorithm}{Algorithms}

\usepackage{tikz}

\usepackage[edges]{forest}
\usepackage[newfloat,frozencache,cachedir=.]{minted}
\usepackage{cite}
\usepackage{xspace}
\usepackage{comment}
\usepackage{multirow}

\usepackage{booktabs, makecell}

\usepackage{threeparttable}

\renewcommand{\paragraph}[1]{\noindent\textbf{#1. }}
\usepackage[framemethod=TikZ]{mdframed}
\usepackage{multirow}
\usepackage[subdued,defaultmathsizes]{mathastext}
\MTnonlettersobeymathxx     % math alphabets will act on (, ), [, ], etc...
\MTexplicitbracesobeymathxx % math alphabets will act on \{ and \}
\MTfamily{\ttdefault}      % we will declare a math version using tt font

\usepackage{makecell}
\usepackage{boldline}
\DeclareUnicodeCharacter{2212}{-}

\newif\ifdraft % set to false to remove comments and todos
\draftfalse
\newcommand{\weon}[1]{\ifdraft{{\footnotesize\color{blue}[Weon: #1]}}\fi}
\newcommand{\mengjia}[1]{\ifdraft{{\footnotesize\color{red}[Mengjia: #1]}}\fi}

\newcommand{\pwd}[1]{\ifdraft{{\footnotesize\color{orange}[Peter: #1]}}\fi}

\newcommand{\rev}[1]{{\color{black}{#1}}}
\newcommand{\del}[1]{}

\newcommand{\attackname}{Speculative Shield Bypass\xspace}
\newcommand{\Ssb}{Speculative shield bypass\xspace}
\newcommand{\ssb}{speculative shield bypass\xspace}
\newcommand{\shortssb}{SSB\xspace}
\newcommand{\Graph}{Speculative information flow graph\xspace}
\newcommand{\graph}{speculative information flow graph\xspace}
\newcommand{\shortgraph}{SIF graph\xspace}

\newcounter{takeaway}
\newcommand{\takeaway}[1]{\vspace{1ex}\noindent\fcolorbox{black}{black!10}{\parbox{\dimexpr\linewidth-2\fboxsep-2\fboxrule\relax}{\refstepcounter{takeaway}\textbf{Takeaway \arabic{takeaway}:} #1}}\vspace{1ex}}

\usepackage{listings}
\input{lstarmlang.sty}
\definecolor{codegreen}{rgb}{0,0.6,0}
\definecolor{codegray}{rgb}{0.5,0.5,0.5}
\definecolor{codepurple}{rgb}{0.58,0,0.82}
\definecolor{backcolour}{rgb}{0.956862745098039,0.956862745098039,0.956862745098039}
\lstdefinestyle{customasm}{
    language=[ARM]Assembler,
    backgroundcolor=\color{backcolour},
    commentstyle=\color{codegreen},
    keywordstyle=\color{magenta},
    numberstyle=\tiny\color{codegray},
    stringstyle=\color{codepurple},
    basicstyle=\ttfamily\small,
    breakatwhitespace=false,         
    breaklines=true,                 
    keepspaces=true,                 
    numbers=left,       
    numbersep=5pt,                  
    showspaces=false,                
    showstringspaces=false,
    showtabs=false,                  
    tabsize=2,
    frame=single, 
    frameround=tttt,
    stepnumber=1,
    numberstyle=\tiny\color{gray}, 
    columns = fullflexible,
    numbersep=5pt,
    xrightmargin=.1in,
    xleftmargin=.1in,
    morekeywords = {memcpy,strlen,call,ret},
}
\lstdefinestyle{customc}{
    language=C,
    backgroundcolor=\color{backcolour},
    commentstyle=\color{codegreen},
    keywordstyle=\color{magenta},
    numberstyle=\tiny\color{codegray},
    stringstyle=\color{codepurple},
    basicstyle=\ttfamily\small,
    breakatwhitespace=false,         
    breaklines=true,                 
    keepspaces=true,                 
    numbers=left,       
    numbersep=5pt,                  
    showspaces=false,                
    showstringspaces=false,
    showtabs=false,                  
    tabsize=2,
    frame=single, 
    frameround=tttt,
    stepnumber=1,
    numberstyle=\tiny\color{gray}, 
    columns = fullflexible,
    numbersep=5pt,
    xrightmargin=.1in,
    xleftmargin=.1in,
    morekeywords = {memcpy,strlen,call},
}
\lstdefinestyle{custompy}{
    language=Python,
    backgroundcolor=\color{backcolour},
    commentstyle=\color{codegreen},
    keywordstyle=\color{magenta},
    numberstyle=\tiny\color{codegray},
    stringstyle=\color{codepurple},
    basicstyle=\ttfamily\small,
    breakatwhitespace=false,         
    breaklines=true,                 
    keepspaces=true,                 
    numbers=left,       
    numbersep=5pt,                  
    showspaces=false,                
    showstringspaces=false,
    showtabs=false,                  
    tabsize=2,
    frame=single, 
    frameround=tttt,
    stepnumber=1,
    numberstyle=\tiny\color{gray}, 
    columns = fullflexible,
    numbersep=5pt,
    xrightmargin=.05in,
    morekeywords = {memcpy,strlen,call, func, uarch, ld},
}

\usepackage{enumitem}
\setlist[itemize]{leftmargin=*, itemsep=0pt}

\newlength{\mysep}
\begin{document}

\title{\LARGE Penetrating Shields: A Systematic Analysis of \\
Memory Corruption Mitigations in the Spectre Era} 
\author{\\ Weon Taek Na$^1$, \space \space Joel S. Emer${^1}{^,}{^2}$,  \space \space Mengjia Yan$^1$\\ \\ 
$^1$MIT, $^2$NVIDIA \\
\{weontaek, jsemer, mengjiay\}@mit.edu
}

\date{}
\maketitle

\thispagestyle{empty}

\begin{abstract}
\vspace{+1ex}

This paper provides the first systematic analysis of a synergistic threat model encompassing memory corruption vulnerabilities and microarchitectural side-channel vulnerabilities.
We study speculative shield bypass attacks that leverage speculative execution attacks to leak secrets that are critical to the security of memory corruption mitigations (i.e., the shields), and then use the leaked secrets to bypass the mitigation mechanisms and successfully conduct memory corruption exploits, such as control-flow hijacking.

We start by systematizing a taxonomy of the state-of-the-art memory corruption mitigations focusing on hardware-software co-design solutions.
The taxonomy helps us to identify 10 likely vulnerable defense schemes out of 20 schemes that we analyze.
Next, we develop a graph-based model to analyze the 10 likely vulnerable defenses and reason about possible countermeasures.
Finally, we present three proof-of-concept attacks targeting an already-deployed mitigation mechanism and two state-of-the-art academic proposals.

\end{abstract}

\section{Introduction}
%\vspace{+2ex}

Memory corruption bugs~\cite{szekeres2013sok, burow2017control, cowan2000buffer}  are one of the oldest security problems in computer systems. 
According to the MITRE 2021 rankings~\cite{mitre}, the top 10 most dangerous software weaknesses encompass multiple memory corruption bugs, including buffer overflows and use-after-frees.

Designing effective and low-overhead  memory safety mechanisms has become a rich research area~\cite{szekeres2013sok} and extensive progress has been made in both academia and industry.
For example, in academia, there have been a vast number of works proposing to leverage architectural innovations or insights to reduce performance and storage overhead of software-only defenses, 
including CHERI~\cite{cheri}, Hardbound~\cite{hardbound}, WatchDog~\cite{watchdog}, WatchdogLite~\cite{watchdoglite}, CHEx86~\cite{chex86}, No-FAT~\cite{no-fat}, ZERO~\cite{zero}, REST~\cite{rest}, Califorms~\cite{califorms}, AOS~\cite{aos},  C3~\cite{c3}, and Morpheus~\cite{morpheus}.
Moreover, the industry has been actively adopting mitigation mechanisms. 
For example, major processor vendors have announced and shipped products with many features supporting increased memory safety, such as Intel MPX~\cite{intelmpx}, Intel CET~\cite{intelcet}, ARM Pointer Authentication~\cite{armpa}, ARM MTE~\cite{armmte}, and SPARC ADI~\cite{sparc_adi}.

Despite the vast number of solutions that have been proposed and deployed, memory corruption attacks continue to pose a serious security threat.
More concerningly, we observe that there is a growing trend of leveraging \emph{synergistic} attack strategies to bypass memory corruption mitigations. %\pwd{Wouldn't italicize this}.
Specifically, instead of only exploiting memory corruption vulnerabilities, adversaries have started to build advanced attacks by \emph{combining} attack strategies which fall into traditionally disjoint threat models, targeting different layers of the system stack.
Notably, multiple recent works have demonstrated that microarchitectural side channels, which exploit \emph{hardware} vulnerabilities, can be used to break Address Space Layout Randomization (ASLR)~\cite{speculativeprobing, hund13, gras17} and ARM Pointer Authentication (PA)~\cite{pacman}, two primary defenses in modern systems for protecting against \emph{software} vulnerabilities.

\paragraph{Motivation: \attackname Attacks} 
We call an attack a \emph{\ssb} attack (SSB attack for short) if it exploits speculative execution to bypass memory corruption mitigations (i.e., the shields), such as ASLR and ARM PA. 
Specifically, a \ssb attack exploits side-channels to leak some secret that is critical to the security of a memory corruption mitigation. Once a speculative shield bypass attack acquires the secret knowledge necessary to bypass the mitigation mechanisms, it exploits this knowledge to conduct a memory corruption exploit, such as control-flow hijacking.%~\cite{burow2017control, roemer2012rop, bletsch2011jop}.

It is critically important to study \ssb attacks.
On one hand, \shortssb attacks out-perform traditional memory corruption attacks, violating both confidentiality and integrity of the system in ways that were not traditionally possible.
On the other hand, \shortssb attacks are easy to miss and difficult to mitigate.
Modern systems have a large and complex attack surface.
To handle such non-trivial complexity, security researchers today often partition the problem space into disjoint threat models, exploring each threat model separately. 
However, to mitigate against \shortssb attacks, researchers must first acquire domain expertise that spans both side-channel vulnerabilities as well as memory corruption vulnerabilities.

\paragraph{This Paper}
We strive to provide the \emph{first} systematic analysis of a threat model that encompasses both memory corruption and microarchitectural side-channel threat models, focusing on \ssb attacks. In this quest, we aim to answer the following questions.

\begin{itemize}
    \item Are any memory corruption mitigations vulnerable to \ssb attacks? 
    How do we identify and classify the insecure mitigations? % 
    
    %\item How does a realistic speculative shield bypass attack exploitation operate?
    %Are there common recipes among such \ssb attacks?
    
    \item What new insights can we learn from the analysis? Can these insights be used to guide the design of future memory corruption mitigation mechanisms to be resilient against \ssb  attacks? %\pwd{First sentence of this bullet is redundant}
\end{itemize}

To answer the above questions, we first systematize a taxonomy 
which allows one to reason about the security of memory corruption mitigations against \ssb attacks.
% \mengjia{I think it matters to talk about spoofable security checks versus unspoofable. that is our new contributions. and also emphasize, no one has studied this unique angle of classification before.}
Through this process, we identify two classes of defenses that are likely vulnerable to \ssb attacks, namely ones that utilize tamperable metadata or randomize the address layout. We find that 10 mitigations fall in these two classes of defenses.

Next, we develop a graph-based model to systematically analyze the 10 likely vulnerable defenses and to reason about possible countermeasures. We call our graphs ``\graph{s}''  (\shortgraph{s} for short). \shortgraph{s} help us precisely visualize the information flow between the security checks imposed by the memory corruption defenses and observable microarchitectural events, identifying sources of microarchitectural side-channel leakage.
%\mengjia{ sources? what about something like patterns?}

To further support our analysis results, we demonstrate proof-of-concept attacks breaking three state-of-the-art defenses, Stack Smashing Protection~\cite{stackguard_canaries}, Always On Heap Memory Safety (AOS)~\cite{aos}, and Cryptographic Capability Computing (C3)~\cite{c3}.

\paragraph{Contributions} 
In summary, we make the following contributions:
\begin{itemize}
    %\item A taxonomy which allows one to reason about the security of memory corruption defenses against \ssb attacks.
    \item A taxonomy identifying the critical characteristics that lead to vulnerabilities in memory corruption mitigations against \ssb attacks.
    \item 
    A graph-based model to analyze and reason about side-channel vulnerabilities in memory corruption mitigations and possible countermeasures. %\mengjia{how about countermeasures?}
    %\item A systematic analysis of 20 defense proposals, among which 10 are vulnerable to either one or both of the two attack vectors. 
    \item Three proof-of-concept demonstrations of \ssb attacks.
\end{itemize}
\section{Background}
\label{sec:bac}
\vspace{-1ex}
%Synergistic attacks\mengjia{fix name} target vulnerabilities across multiple layers of the system stack to achieve various end goals. 
\Ssb attacks exploit the synergies that arise at the convergence of side-channel vulnerabilities and memory corruption vulnerabilities. 
%In this paper, we focus on the synergies that arise at the convergence of side-channel vulnerabilities and memory corruption vulnerabilities.
In this section, we give a brief background on these vulnerabilities and their exploits.

\subsection{Micro-architectural Side Channel Attacks}
\label{sec:bac_side_channel}
\vspace{-1ex}

Micro-architectural side-channels are a class of vulnerabilities that enables an attacker to \textit{leak information} and steal some \textit{secret} from the victim application running on the same machine. 
To do so, the attacker monitors the side effects of the victim’s actions on various micro-architectural structures.
As formalized in \cite{dawg}, a micro-architectural side-channel attack comprises of either pre-existing or attacker-generated code run in the victim’s \textit{security domain} that 1) accesses secret information and 2) transmits that information over a communication channel that 3) is received by an attacker.
Despite existing protection and isolation mechanisms, the signal transmitted over the channel leaks a secret that should be confined to its security domain. 
From a telecommunications perspective, the \textit{transmitter} is in the victim’s code, and the \textit{receiver} is in the attacker’s code, and the medium of the \textit{communication channel} is the micro-architectural state that can be modulated by the activity of the transmitter~\cite{casa}. In general, the receiver measures the microarchitectural resource usage of the victim, using explicit timer instructions~\cite{flushreload, primeprobe}, or a custom timer constructed through shared memory~\cite{dutta2021leaky}. 
Side-channel attacks have been demonstrated on an ever-growing list of microarchitectural structures~\cite{szefer2019survey}, such as
branch predictors~\cite{aciiccmez2007predicting, evtyushkin2018branchscope}, caches~\cite{flushreload, primeprobe, primeabort, flushflush, neve2006advances,  deng2020benchmark, bonneau2006cache, lipp2016armageddon, gullasch2011cache},
translation lookaside buffers~\cite{tlbbleed},
on-chip networks~\cite{wassel2013surfnoc}, 
and memory controllers~\cite{wang2014timing}. 
%Among the abundant side-channels in today's processors, caches are arguably the most commonly exploited. 
In this paper, our proof-of-concept attacks exploit cache-based side-channels. 

The most popular variants of cache-based side-channel attacks are Flush+Reload~\cite{flushreload} and Prime+Probe~\cite{primeprobe}.
In Flush+Reload, the attacker and the victim share memory.
%The attacker first flushes one of the victim's cache line that he or she wants to monitor, out of the cache.
The attacker first flushes one of the victim's cache lines out of the cache, and 
then waits for the victim to access the cache line, causing modulation on the cache line (i.e., changing the state of the cache line). 
%The modulation is detected by the attacker, by measuring the time of accessing the victim address. 
The attacker then detects the modulation by measuring the time to access the same cache line. 
%If the access time of the victim address is low (i.e., cache hit), the attacker can infer that the cache line was accessed by the victim.
If the access time is short (i.e., cache hit), the attacker can infer that the cache line was accessed by the victim.

In Prime+Probe, the attacker first constructs eviction sets. An eviction set is a group of memory addresses that map to the same cache set with at least as many lines as the associativity of the cache. 
%An eviction sets is used by the receiver to monitor the cache lines in a cache set. 
%Once the attacker successfully constructs the necessary eviction sets, the attack repeats the following steps. 
The attacker then repeats the following steps. 
First, the attacker populates cache sets that he or she wants to monitor, with the eviction sets. 
Next, the attacker waits for the victim to access some cache lines. 
Finally, the attacker detects the modulation of the cache by measuring the time to re-access the eviction sets. 
If the re-access time of an eviction sets is high (i.e., cache miss), the attacker can infer that the corresponding cache set was accessed by the victim.

%\mengjia{optional: we can end this paragraph with a forward pointer by saying that we use prime+probe in our poc in section xx.}

\subsection{Speculative Execution Attacks}
\label{sec:bac_spec_exec}
\vspace{-1ex}

Speculative execution attacks are a subset of microarchitectural side-channel attacks that exploit the speculative nature of modern processors.
Modern processors execute instructions ahead-of-schedule, by predicting the outcome of control-flow decisions and \textit{speculatively executing} instructions based on those predictions. 
%Due to the nature of speculation, the instruction may turn out to be valid or invalid. Later, if an instruction turns out to be valid, the instruction is retired, and the processor's architectural state before the speculation is discarded. 
If a prediction turns out to be correct, the speculatively executed instructions are retired.
Otherwise, the instructions are squashed and the processor's state is rolled back to the architectural state before these instructions.
%However, the micro-architectural state is modified as result of the incorrect speculation, causing side-channels to be modulated, which may allow secrets to leak. 
%In other words, these mis-speculated instructions are executed \textit{transiently} (first, they are executed, but later they vanish). Hence, speculative execution attacks are also called \textit{transient execution} attacks.~\cite{canella2019systematic}. 
These mis-speculated instructions which are deemed to be squashed are called \emph{transient instructions}~\cite{canella2019systematic}. 
The micro-architectural state that is modified by these transient instructions causes modulations to the  side-channels, ultimately leaking secrets. 

%\mengjia{we should also mention that speculative execution attacks are sometimes refered to as "transient execution attacks" and cite~\cite{canella2019systematic}. Also precisely define what are transient instructions: instructions that are speculatively executed but deemed to be squashed.}

By exploiting mis-speculated execution, an attacker can circumvent software invariants, exercising code paths that should be unreachable, and leak \textit{any} secret in the victim's addressable virtual memory.
%\mengjia{there is no clear definition of access and transmit code. For people who are not familiar with these terminologies, they will be confused and it does not help how exactly spectre and metldown works.
%Another thing is that I do not understand is that there exist many differences between spectre and meltdown, but you seem to try to emphasize one uses victim's existing code and the other uses attacker's code.
%Why we try to emphasize this point for our paper?} 
%Spectre~\cite{spectre} and Meltdown~\cite{meltdown} first discovered this phenomena. As discussed in \cite{dawg},  Spectre synthesizes an access and transmitter code out of existing code in the victim, by coercing branch predictor state to encourage mis-speculation along an attacker-selected code path. Meltdown explicitly programs the access and transmitter code that speculatively execute the access and transmitter code and illegally leak the secret.  The illegal access causes an exception to be raised, but the secret is transmitted via micro-architectural side effects.
Today, a plethora of speculative execution attacks have been demonstrated, including Spectre~\cite{spectre}, Meltdown~\cite{meltdown} and their many variants~\cite{horn2018speculative, koruyeh2018spectre, maisuradze2018ret2spec,  stecklina2018lazyfp, vanbulck2018foreshadow, weisse2018foreshadow}. %\mengjia{more citations?}

\subsection{Memory Corruption Attacks}
\label{sec:bac_memattack}
\vspace{-1ex}

%Memory corruption bugs are an orthogonal class of vulnerabilities with a long history of exploits. 
Memory corruption bugs are an orthogonal class of vulnerabilities to microarchitectural side channels.
In addition to leaking secrets, memory corruption bugs can also be exploited to violate integrity, compromise authority and perform  arbitrary code execution.
%gain authority and arbitrary code execution.\mengjia{I added this to say it is different from side channel.. @weon check}
Software written in low-level languages like C or C++ are prone to memory corruption bugs. 
Data from Google and Microsoft~\cite{google-bugs, microsoft-bugs} indicate that nearly 70\% of the security bugs found today are due to memory corruption vulnerabilities in C/C++ software.
%\mengjia{comment out a sentence to save space... }
%\weon{Put it back in. feels a little awkward without it. hopefully we can take sth else out}

The most prevalent memory corruption vulnerabilities violate either spatial or temporal safety. Spatial vulnerabilities include buffer overflow and underflow vulnerabilities, while temporal vulnerabilities are largely dominated by use-after-free vulnerabilities. 
Spatial safety violations can be subdivided further into adjacent and non-adjacent overflows~\cite{nonadjacent-buffer-overflow}.
% Starting from an initial allocation, adjacent buffer overflows are limited to adjoining allocations, whereas non-adjacent buffer overflows are able to skip over the adjoining allocations to perform arbitrary reads or possibly writes.
%\mengjia{I commented out the last sentence that explains adjacent/non-adjacent overflow, since it is a bit difficult to follow without an example.}

We provide examples of three types of memory corruption vulnerabilities in \cref{lst:mem-corrupt}.
%adjacent and non-adjacent buffer overflows and use-after-free vulnerabilities below 
%\pwd{Don't forget the figure title}. 
%An attacker can exploit the adjacent buffer overflow vulnerability in line 4, to read in more than the size of the \texttt{buf}, and thereby overwrite locations adjacent to \texttt{buf}.
An attacker can exploit the adjacent buffer overflow vulnerability in line 4 by passing in a string longer than the size of \texttt{buf}, and thereby overwrite locations adjacent to \texttt{buf}.
In line 7, an attacker can exploit the non-adjacent buffer overflow vulnerability by controlling the arguments \texttt{x} and \texttt{y} to perform an arbitrary write. In line 10, an attacker can exploit the use-after-free vulnerability to overwrite a freed location.
%\mengjia{can we cite the slides from microsoft for these examples, especially the non-adjacent one...}
%\weon{I added a citation a paragraph above}
%\mengjia{for use-after-free, 1. since we do not care about adjacent or not, shall we use 0 as the index, instead of "x"?; 2. it is a read, why not make it a write?}
%\weon{was trying to show a diversity. so tried having a mixture of read andd write. changed it to write. changed to 0 for index.}

\vspace{-2ex}
\begin{listing}[h]
\begin{minted}[frame=single,fontsize=\footnotesize,      obeytabs=true,tabsize=4,numbersep=3pt,linenos,escapeinside=||]{c}
void 
vulnerable(int x, int y) { // x, y are attacker controlled
  char buf[128];
  gets(buf); // 1. adjacent overflow vulnerability 

  int* alloc1 = (int* ) malloc (N * sizeof(int));
  alloc1[x] = y; // 2. non-adjacent overflow vulnerability
  
  free(alloc1);
  alloc1[0] = y; // 3. use-after-free vulnerability
}
\end{minted}
\vspace{-4ex}
\caption{Examples of memory corruption vulnerabilities.}
\label{lst:mem-corrupt}
\end{listing}
\vspace{-2ex}

Once attackers have access to such memory corruption vulnerabilities, they can target different program properties to achieve various end goals.

%\mengjia{I suggest to organize these attacks slightly differently. Let me know how you think about it.
%Because the second category "Control-Flow Hijacking Attacks" is mostly studied and a lot of mitigations target it. we should spend most text on it.
%Then the other types should be mentioned together in a single paragraph: say there are other attacks , such as ....., and keep the discussion brief.
%The goal is to get people something concrete to think about ... Not to be as comprehensive as possible.}

\paragraph{Control-Flow Hijacking Attacks} Corrupting a code pointer causes a control-flow transfer to anywhere in executable memory. Code pointers include return addresses on the stack and function pointers anywhere in memory. 
%In the early days of the attack on memory, it was common for an attacker to include the malicious code inside the payload and overflow a buffer on the stack. 
%This allowed the attacker to point the overwritten return address to a known address on the stack, where the malicious code resides~\cite{phrack-stacksmashing}. 
%Such code injection attacks are no longer practical on all modern systems due to the deployment of non-executable memory, which marks data pages as non-executable. 
Today, control-flow hijacking attacks are executed using attack techniques called return oriented programming (ROP)~\cite{roemer2012rop} which corrupt return addresses or jump-oriented programming (JOP)~\cite{bletsch2011jop} which corrupt indirect code addresses (typically function pointers). Collectively, ROP and JOP are called code reuse attacks (CRAs). 
To mount a CRA, an attacker first analyzes the victim's code to identify sequences of
instructions that end with a return or jump instruction (called \textit{gadgets}). 
Next, the attacker uses a memory corruption vulnerability to inject a sequence of target addresses corresponding to a sequence of gadgets. 
Later, when a code pointer is de-referenced or the function returns, it moves to the
location of the first gadget. At the termination of the first gadget, the control-flow is transfered to the second gadget (and so on) by a control flow instruction such as \texttt{return} or \texttt{jump}.
%\mengjia{I hope to see the phrase "*chain* gadgets together ...."}

\paragraph{Other Attacks} 
%Traditionally, it was common for an attacker to overwrite program code directly with adversarial payload to perform \textit{code corruption attacks}~\cite{citations?}. \mengjia{need citation here}
%or inject the adversiral payload on the stack to perform \textit{code injection attacks}.
%\mengjia{I want to remove code injection, because it is one type of control-flow hijack @weon check..} 
%\pwd{Code injection is a control-flow hijacking attack, but 
%Today, code corruption attacks are no longer feasible as hardware and software transparently mark code pages as non-writable and data pages as non-executable. 
%\mengjia{JIT?} \weon{JIT would make code corruption attacks feasible, yes}
\textit{Data-oriented programming} attacks~\cite{hu2016data, ispoglou2018block} cause malicious end results without changing the control flow of the program, by only manipulating  data pointers.
% Prior works have demonstrated attacks achieving arbitrary computations on program input~\cite{hu2016data, ispoglou2018block}, by manipulating only data pointers. 
\textit{Data corruption attacks}~\cite{chen2005non} target non-pointer data stored in memory. 
For example, an attacker can bypass authorization checks by manipulating program flags.
%\mengjia{+ in this paper, our synergistic attacks aim to conduct CRA after bypassing mitigation mechanisms}
%\weon{i think we make it clear enough now.}

%\paragraph{Code Corruption Attacks}  Traditionally, it was common for an attacker overwrite program code directly with adversarial payload to perform. Today, code corruption attacks are no longer feasible as hardware and software transparently mark code pages as non-writable.

%\paragraph{Data-Flow Hijacking Attacks} Data-oriented programming (DOP) attacks
% cause malicious end results without changing the control
%flow of the program. 
%Prior works have demonstrated attackers achieving arbitrary computations on program input~\cite{hu2016data, ispoglou2018block}, by manipulating only data pointers.

%\paragraph{Data Corruption Attacks} This last class of attacks targets
%non-pointer data stored in memory. 
%Prior work has demonstrated an attacker bypassing selective checks by manipulating program flags~\cite{chen2005non}.

%\include{sections/bac-history-memorycorruption}

%\begin{verbatim}
%    - Recently, the kernel, which is not crash-resistant, has been also shown to leak the locations of the code when crashing \textit{transiently}. crashes which  side-channels~\cite{speculativeprobing}. 
%    - code integrity enforcing fails for JIT
%\end{verbatim}

\section{Threat Model}
\label{sec:threatmodel}

The threat model of \ssb attacks considers two popular scenarios:
(1) A sandbox scenario where an attacker is confined to a sandboxed environment with limited code execution on the target machine, such as in Javascript sandbox inside a browser, Google Native Client~\cite{nacl}, and Linux eBPF~\cite{ebpf}.
In the sandbox scenario, the host (sandbox creator) is the victim.
(2) A cloud scenario where an attacker runs on a remote device and interacts with another program, such as in remote web servers and SGX enclaves~\cite{costan2016intel}. In the cloud scenario, the remote web server and the remote SGX enclave are the victim.
We make the following assumptions.

First, there exists an exploitable memory corruption vulnerability in the victim, which allows the attacker to write to some memory locations in the victim program.
The attacker’s goal is to exploit this vulnerability to perform a memory corruption exploit, such as control-flow hijacking.

Second, to defend against software vulnerabilities, the victim incorporates one of the respective memory corruption mitigation considered in this paper.
Hence, the attacker's intermediate goal is to bypass this defense (i.e., the shield).

Third, the attacker is able to perform a micro-architectural side-channel attack on the machine running the victim program. 
Thus, if the defense mechanism that the victim incorporates relies on the confidentiality of some secret information, it may be vulnerable to a \ssb attack.

% Thus, the defense mechanism that the victim incorporates (which possibly relies on the confidentiality of some secret information), is vulnerable to information leaks.

%to leak secrets that are potentially critical to the security of the mitigation mechanism.. 
% Thus, the defense mechanism that the victim incorporates (which possibly relies on the confidentiality of some secret information), is vulnerable to information leaks.
%\mengjia{the old last sentence is a bit forward looking. we do not know this before we do the taxonomy right? I did some edits.}

\section{Speculative Shield Bypass Attacks}
\label{sec:ssb}

A \ssb attack (\shortssb attack for short) consists of two critical steps: 1) performing an \textit{information disclosure attack} using a side-channel vulnerability to gain some secret that is critical to a memory corruption defense, and 
2) using the leaked secret knowledge to \textit{spoof the security check} of the defense while performing a memory corruption attack.

In this section, we study an example of a \ssb attack called Speculative Probing~\cite{speculativeprobing}. 
Speculative Probing is a state-of-the-art \shortssb attack penetrating ASLR.
ASLR adds random offsets to memory segments, such as the code segment, the stack, and the heap, to obfuscate the locations of code-reuse gadgets which are necessary to construct code-reuse attacks~\cite{pax_aslr, usenix_aslr}.
As a result, attackers need to correctly guess the secret offsets to calculate the addresses of the gadgets.

In Speculative Probing, the attacker first exploits a side-channel vulnerability to leak the secret offset, by targeting the following code snippet in the vicim:
\verb|if (cond) { call f_ptr(); }|.
Specifically, the attacker exploits a memory corruption vulnerability (\cref{sec:bac_memattack}) to overwrite \texttt{f\_ptr} using a guessed offset.
To check whether the offset is guessed correctly, the attacker triggers mis-speculation on the branch to execute the indirect call. 
During the indirect call, the memory management unit (MMU) performs a security check on the address added with guessed offset.
If the offset is correct, the guessed address is mapped, and hence the security check passes, modulating the memory system. 
Otherwise, the security check triggers a speculative exception (which is squashed) instead.
Thus, the attacker can monitor whether the guessed address is mapped or not using microarchitectural side-channels.
The attacker repeats the above process, abusing the security check performed by the MMU as a side-channel, ultimately brute-forcing the secret offset.
% Once the secret offset is leaked, the attacker proceeds to the second step of conducting a memory corruption exploit, by setting the condition \texttt{cond} to true, triggering a traditional control-flow hijacking attack to the disclosed gadget address.
Once the secret offset is leaked, the attacker proceeds to the second step of conducting a memory corruption exploit, by setting the condition \texttt{cond} to true.
This time, the indirect call will commit, rather than being squashed, resulting in a traditional control-flow hijacking attack to the disclosed gadget address.

To this date, there already exist two case studies demonstrating \ssb attacks, namely Speculative Probing~\cite{speculativeprobing} breaking ASLR, and PACMAN~\cite{pacman} breaking ARM Pointer Authentication.
% Yet, most memory corruption mitigation proposals from both industry~\cite{intelmpx, armmte,intelcet,armbti, armpa} and academia~\cite{hardbound, watchdog, watchdoglite, aos, chex86, califorms, rest, morpheus, cheri, no-fat,zero,c3} continue to consider side channels out-of-scope of their threat model. 
Yet, most memory corruption mitigation proposals from both industry and academia continue to consider side channels out-of-scope of their threat model.
We reason that this trend may incur a critical security crisis if memory corruption mitigations continue to be deployed with such a narrow threat model in consideration, despite being vulnerable to \shortssb attacks in the wild. 

To better anticipate this crisis, we develop a taxonomy (\cref{sec:taxonomy}) to identify the critical characteristics that lead to \ssb vulnerabilities  and a graph-based model (\cref{sec:model}) to analyze and reason about side-channel vulnerabilities in these memory corruption mitigations.

\begin{figure*}[t]
\centering
\begin{tikzpicture}[
every node/.style = {draw=black, 
	inner color=gray!5,
	outer color=gray!10,
	fill = yellow,
	rounded corners, 
	thick,
	text width=49mm, 
	minimum width=1cm,
	%rounded corners=3,
	%text height=1.5ex,
	%text depth=0ex,
	align=center,
	anchor=north, 
	font=\footnotesize},
level distance = 6mm,
sibling distance = 82mm,
%edge from parent fork down
]
\node {Memory Corruption Mitigations}
child{ node {\textcolor{red}{Spoofable Security Checks}}[sibling distance = 54mm]
    child{ node {{\textcolor{red}{Address Layout Randomization}}}
    	child{ node {
    	{\textcolor{red}{ASLR}}~\cite{pax_aslr, usenix_aslr},\\ {\textcolor{red}{Califorms (no-adjacent)$\dagger$}}~\cite{califorms},\\ {\textcolor{red}{Morpheus}}~\cite{morpheus} }}
    }
    child{ node {{\textcolor{red}{Tamperable Metadata Augmentation}}}
    	child{ node {
    	{\textcolor{red}{Stack Smashing Protection}}~\cite{stackguard_canaries}, 
    	{\textcolor{red}{ARM PA}}~\cite{qualcomm_white_paper}, 
    	{\textcolor{red}{ARM MTE}}~\cite{armmte}, 
    	{\textcolor{red}{SPARC ADI}}~\cite{sparc_adi}, 
    	{\textcolor{red}{AOS}}~\cite{aos}, 
    	{\textcolor{red}{C3}}~\cite{c3}, {\textcolor{red}{No-FAT (temporal)$\ddagger$}}~\cite{no-fat} }}
    }
}
child{ node {Unspoofable Security Checks}
child{ node {Tamperproof Metadata Augmentation}
    child{ node {
	CHERI~\cite{cheri}, Intel MPX~\cite{intelmpx}, Hardbound~\cite{hardbound}, WatchDog~\cite{watchdog},
	WatchdogLite~\cite{watchdoglite}, CHEx86~\cite{chex86},    	Intel CET~\cite{intelcet}, ARM BTI~\cite{armbti}, ZERO~\cite{zero}, REST~\cite{rest},
    Califorms (adjacent)$\dagger$~\cite{califorms}, No-FAT (spatial)$\ddagger$~\cite{no-fat} }}
}
}
;
\end{tikzpicture}
\\

    \footnotesize \textsuperscript{$\dagger$}Califorms maintains tamperproof metadata for adjacent vulnerability protection and \\adds randomness to the size of the metadata for non-adjacent vulnerability protection.\\
    \footnotesize \textsuperscript{$\ddagger$}No-FAT maintains tamperproof metadata for spatial protection, but tamperable metadata for temporal protection.
	\caption{Taxonomy of proposed or deployed memory corruption mitigations when focusing on \ssb attacks. Mitigations in {\textcolor{red}{red}} are likely vulnerable to \ssb attacks.}
	%\mengjia{1. move the comments to footnote? otherwise taking too much space?} 
% 	\weon{added adi.}
% 	\mengjia{can we remove Califorms from the unspoofable to avoid confusion? Also add temporal safety and spatial safety to No-fat to calrify why they appear twice.}
% 	\weon{added adjacent vs non-adjacent for califorms and temporal vs spatial for no-fat. Califorms for adjacent is same as REST which is tamperproof.}
% 	\label{fig:taxonomy}
    \label{fig:taxonomy}
\end{figure*}

\ifdraft
\begin{figure}
    \centering
    \begin{tabular}{@{}p{8.5cm}@{}}
    \includegraphics[width=\columnwidth]{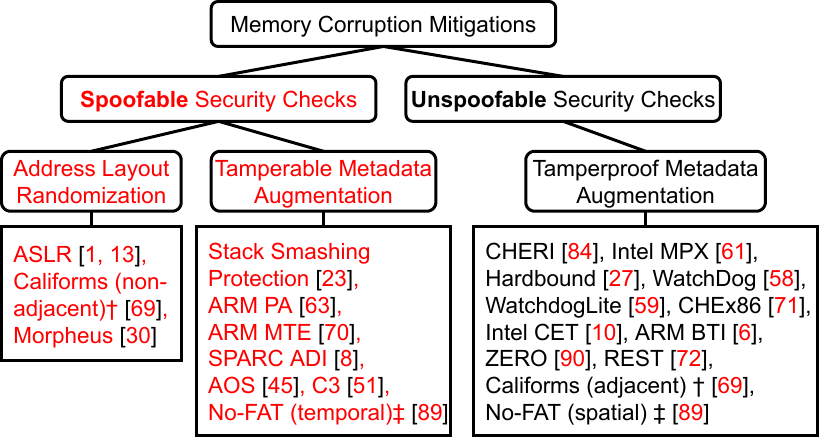}\\
    \footnotesize \textsuperscript{$\dagger$}Califorms maintains tamperproof metadata for adjacent vulnerability protection and adds randomness to the size of the metadata for non-adjacent vulnerability protection.\\
    % ; the randomness regarding the size of the metadata is vulnerable to \ssb attacks.\\
    \footnotesize \textsuperscript{$\ddagger$}No-FAT maintains tamperproof metadata for spatial protection, but tamperable metadata for temporal protection.
    % ; the tamperable metadata is vulnerable to \ssb attacks.
    \end{tabular}
    \caption{Taxonomy of proposed or deployed memory corruption mitigations when focusing on \ssb attacks. Mitigations in {\textcolor{red}{red}} are likely vulnerable to \ssb attacks.}
\end{figure}
\fi

%\section{Systematic Analysis of Memory Corruption Mitigations against Synergistic Attacks}
% \section{Classification for Analyzing SSB Vulnerabilities}
\section{Classifying Mitigations for Speculative Shield Bypass Analysis}
\label{sec:taxonomy}

In this section, we aim to develop a taxonomy which allows one to reason about the security of memory corruption mitigations against \ssb attacks. 
%Our ultimate goal is to identify and analyze a meaningful subset of common weaknesses in memory corruption mitigations that may be exploited due to the presence of side-channel vulnerabilities in modern processors.
Our ultimate goal is to identify and analyze the critical characteristics shared among a subset of memory corruption mitigations, that may be exploited using side-channel attacks.
% due to the presence of side-channel vulnerabilities in modern processors.
%\mengjia{subset of weakness or subset of mitigation? an attempt: identify what critical characteristics make a scheme vulnerable to SSB?}

We start by assessing the fundamental roots of current taxonomies of memory corruption defenses and why they are insufficient in analyzing these defenses against \ssb attacks.
Fundamentally, memory corruption vulnerabilities occur due to the lack of sufficient security checks in the vulnerable programs.
Memory corruption defenses aim to protect these possibly vulnerable programs by augmenting them with automated security checks during their runtime. 
%As such, past taxonomies have tended to first develop comprehensive models of memory corruption attacks, and then categorize the defenses inside these attack models in relation to which steps of the attack the augmented security checks occur. 
As such, past taxonomies~\cite{szekeres2013sok, novkovic2021taxonomy,saito2016survey, cowan2000buffer, burow2017control} have tended to categorize the defenses inside memory corruption attack models, in relation to which steps of the attack the augmented security checks occur. 
While this approach is helpful in understanding which types of memory corruption attacks can be mitigated by the varying security checks imposed by the respective defenses, it is not as useful for understanding the resiliency of the defenses against \ssb attacks.

Recall from \cref{sec:ssb},
%\mengjia{TODO: check after a full pass of the paper that we do define SSB in S4?} 
% \ssb attacks 1) first perform an information disclosure attack using a side-channel vulnerability to gain some secret knowledge and 2) then uses the leaked secret knowledge to spoof the security check of a defense.
%\pwd{Did you guys consider using other words instead of spoof (e.g. forge)? The grammar with spoof is a little weird (i.e. you're spoof the metadata, you don't spoof the security check)} the security check of a defense.
% using a memory corruption vulnerability. 
the second step in a \ssb attack is to make use of some leaked secret to spoof the security check of a defense.
%\mengjia{made a change here. check?}
In other words, from a \shortssb attack perspective, it is not a question of which step of a memory corruption attack that a security check occurs in. 
Rather, the question is \emph{whether an attacker can forge rogue information that can spoof a security check}.
Our taxonomy primarily focuses on answering this question.
% , since this ultimately determines whether there is any incentive for an attacker to perform an information disclosure attack at all. 

We show our taxonomy in \cref{fig:taxonomy}  with examples of state-of-the-art mitigation mechanisms (focusing on hardware-software co-designs) from both academia\hbox{~\cite{hardbound, watchdog, watchdoglite, aos, chex86, califorms, rest, morpheus, cheri, no-fat,zero,c3, usenix_aslr}} and industry\hbox{~\cite{intelmpx, armmte,intelcet,armbti, armpa, pax_aslr}}.
% for each category.
Mitigations in \textcolor{red}{red} are likely vulnerable to \ssb attacks.
%\mengjia{are or potentially are? because we also color MTE and ADI}
We first divide mitigations depending on whether they perform \emph{spoofable} security checks (i.e., given the ability to corrupt a pointer, an attacker can also corrupt additional input that also goes into the security checks with the pointers) or \emph{unspoofable} security checks (i.e., while a pointer may be corruptible, the additional input to the security check cannot be controlled by an attacker). 
%We first divide mitigations depending on whether they perform \emph{spoofable} security checks or \emph{unspoofable} security checks.
%Given a malicious pointer that has been corrupted by an attacker, spoofable security checks can be tricked to incorrectly classify the pointer as a legal pointer, while unspoofable security checks will always classify it as an illegal pointer.
%\mengjia{add this sentence to clarify}

% For spoofable security check based defenses, we further subdivide them depending on what kind of information is used in the security checks (i.e., what kind of information the attacker aims to leak). 
For spoofable security check based defenses, we further subdivide them depending on what kind of information is used in the security checks, which is also the information that the attacker aims to leak. 
\emph{Address layout randomization} based defenses add random ``\emph{offsets}'' to the address of various (sub-)objects, entangling these offsets with the originally predictable (sub-)object addresses. 
On the other hand, \emph{tamperable metadata augmentation} based defenses augment addresses or pointers with additional information called ``\emph{metadata}'' that is maintained separately from the pointers. 
% Since these defenses are spoofable\mengjia{security checks are spoofable, not defense?} for an attacker who is equipped with the correct offset or metadata, these defenses maintain this offset or metadata 1) as a \emph{secret} private to its security domain and 2) with high entropy. 
% \mengjia{what does private, and security domain refer to here?}
Since the security checks of these defenses are spoofable
%\mengjia{security checks are spoofable, not defense?} 
for an attacker who is equipped with the correct offset or metadata, these defenses maintain this offset or metadata as a \emph{secret} with high entropy so that it is difficult for the attacker to guess it correctly. 
%\mengjia{tried to simplify to just say it is secret.}

For unspoofable security check based defenses, we found that all defenses in this category can also be categorized as \emph{tamperproof metadata augmentation} based defenses which similarly augment addresses or pointers with metadata. 
However, they differ greatly in that they maintain unspoofable security checks by enforcing the integrity of their metadata. In fact, as the security checks are unspoofable, these defenses do not share the incentive of maintaining the privacy or high entropy of their metadata. Thus, the metadata in these defenses are generally public information (i.e., not secret), nullifying any merits of information disclosure.
% We now explain each category in more detail.

We now provide examples for each category to show how spoofable and unspoofable security checks work.
% \mengjia{maybe say we show examples of how security checks can be spoofed in detail and summarize existing techniques to make it not spoofable?}

% \blfootnote{{$\dagger$}Califorms maintains tamperproof metadata for adjacent vulnerability protection and adds randomness to the size of the metadata for non-adjacent vulnerability protection; the randomness regarding the size of the metadata is vulnerable to \ssb attacks.
% }
% \blfootnote{
% {$\ddagger$}No-FAT maintains tamperproof metadata for spatial protection, but tamperable metadata for temporal protection; the tamperable metadata is vulnerable to \ssb attacks.
% }

\paragraph{Address layout randomization} 
Address layout randomization based defenses randomize the position of (sub-)objects in virtual memory. Such defenses include ASLR~\cite{pax_aslr, usenix_aslr}, Morpheus~\cite{morpheus} and Califorms~\cite{califorms}. 
Specifically, these defenses add random offsets to the addresses of various (sub-)objects, entangling the offset with the originally predictable addresses.
Because a memory corruption exploit which overwrites a pointer allows an attacker to control the address (and hence the offset), address layout randomization schemes can be categorized as performing spoofable security checks.
Specifically, we formulate the security check below with attacker-controlled variable in \textcolor{blue}{blue} and secret in \textcolor{red}{red}.
\begin{minted}[fontsize=\small,      obeytabs=true,tabsize=4,numbersep=3pt,escapeinside=||]{python}
if (|\textcolor{blue}{corrupted\_ptr}| = addr+|\textcolor{red}{rand\_offset}|): pass;
else: fail;
\end{minted}
%\pwd{Do you want to convert these to listings to stay consistent?}
%\mengjia{maybe not... need to save space.}
%If the secret offset is leaked, the attacker can easily manipulate the value of \texttt{\small{corrupted\_ptr}} to bypass the security check.
If the secret offset is leaked, an attacker can adjust the value of \texttt{\small{corrupted\_ptr}} appropriately to bypass the security check.
% \mengjia{this sentence is replicated from above.}
% The insight behind this approach is that if the addresses of (sub-)objects in the virtual memory are unknown, an adversary is unable to divert the control flow reliably, and thus can be caught using an illegal address (i.e., wrong offset) during a security check. 
% When in action, security checks are performed for each memory access by first taking the memory address being accessed and checking whether the address is legal (i.e., the added offset is correct). 

% Because a memory corruption exploit which overwrites a pointer allows an attacker to control the address (and hence the offset) being used for the security check, address layout randomization schemes can be categorized as performing spoofable security checks.
% To mitigate possible spoofing attempts from an attacker, address layout randomization schemes maintain the offset as a secret that is private to their security domain and with high entropy to mitigate guessing attempts. \mengjia{repeat..}
% However, if side-channel vulnerabilities can be exploited to leak the secret offset from a victim's security domain to an attacker, then the attacker can next perform a memory corruption attack while leveraging this secret offset to successfully bypass address layout randomization defenses. 
% Examples of address layout randomization schemes include ASLR~\cite{pax_aslr, usenix_aslr}, Morpheus~\cite{morpheus} and Califorms~\cite{califorms}. 

\paragraph{Tamperable Metadata Augmentation} 
Tamperable metadata augmentation based defenses augment addresses or pointers with separately maintained metadata. These metadata are used to perform varying security checks against the addresses or pointers.
%Tamperable metadata augmentation based defenses augment addresses or pointers with separately maintained metadata, which are used to perform varying security checks against the addresses or pointers.
% The insight behind this approach is that if addresses or pointers are augmented with metadata, security checks can be automatically performed during runtime as a function of the metadata with the addresses or pointers.
However, the integrity of the metadata is not enforced in these schemes.
For example, stack smashing protection \cite{stackguard_canaries}
maintains its metadata on the stack.
ARM Pointer Authentication~\cite{qualcomm_white_paper}, ARM MTE~\cite{armmte}, AOS~\cite{aos}, No-FAT (temporal safety)~\cite{no-fat}, and SPARC ADI~\cite{sparc_adi} maintain their metadata inside unused upper bits of pointers. C3~\cite{c3} re-formats pointers to entangle the metadata with the pointers.
%\joel{No-FAT is listed as both tamperable and non-tamperable?}\mengjia{annotate with temporal safety. Should do the same in the taxonomy figure.}
%\mengjia{how about c3?}
% Tamperable metadata augmentation based defenses maintain their metadata where the integrity of the metadata cannot be enforced. 
% For example, \cite{stackguard_canaries}
% maintains its metadata on the stack and \cite{qualcomm_white_paper, armmte, aos, c3, no-fat} maintain their metadata inside unused upper bits of pointers. 
Since these locations can be overwritten by a memory corruption exploit, the security checks can be spoofed.

%We show how to spoof a security check using ARM MTE as an example.
As an example, we formulate the security check of ARM MTE below.
ARM MTE maintains a tag inside unused upper bits of a protected pointer, and a memory tag that is associated with the location addressed by the pointer.
A security check succeeds if the tag inside the pointer matches the memory tag.
%ARM MTE maintains a tag inside unused upper bits of a protected pointer, and a memory tag that associates with the location addressed by the pointer.
\begin{minted}[fontsize=\small,      obeytabs=true,tabsize=4,numbersep=3pt,escapeinside=||]{python}
if (|\textcolor{blue}{corrupted\_tag}| = |\textcolor{red}{memory\_tags}|[|\textcolor{blue}{corrupted\_ptr}|]): pass;
else: fail;
\end{minted}
If the secret memory tag is leaked, the attacker can adjust the value of the tag inside the pointer appropriately to bypass the security check.

% To mitigate possible spoofing attempts, tamperable metadata augmentation based defenses maintain the metadata as a secret that is private to their security domain and with high entropy.
% Thus, if an attacker can leak this secret metadata from the victim's security domain, then the attacker can next perform a memory corruption attack while utilizing this leaked secret metadata to spoof security check(s) performed.
% Examples of such defenses include Stack Smashing Protection~\cite{stackguard_canaries}, ARM Pointer Authentication (PA)~\cite{qualcomm_white_paper}, ARM Memory Tagging Extension (MTE)~\cite{armmte}, Always On Heap Memory Safety (AOS)~\cite{aos}, Cryptographic Capability Computing (C3)~\cite{c3}, and No-FAT~\cite{no-fat}.

\paragraph{Tamperproof Metadata Augmentation}
Tamperproof metadata augmentation based defenses similarly augment addresses or pointers with metadata that is used to perform varying security checks. 
The difference lies in the fact that instead of relying on privacy and high entropy of the metadata, these defenses directly enforce integrity of their metadata.
% by other means. 

Today, there exists three approaches for enforcing metadata integrity. First, mitigations that perform out-of-bound security checks for all memory accesses can place the metadata in an area outside the bounds of any other object in the address space called \emph{shadow memory}~\cite{intelmpx, hardbound, watchdog, watchdoglite, chex86}. Alternatively, mitigations can transparently tag the memory addresses that store the metadata by leveraging additional specialized hardware called \emph{tagged memory}~\cite{cheri, zero, rest}. Lastly, mitigations can leverage \emph{page attributes} to enforce integrity of the pages that hold the metadata, as in \cite{intelcet, armbti}.
Since the metadata which is used to perform security checks cannot be modified by an attacker, the security checks that are performed by these defenses cannot spoofed. As such, these defenses are unspoofable and thus are not vulnerable to \ssb attacks. 

%\mengjia{I remove the examples, since somehow we will not focus on them anyway... so...just remove... you already cited in the previous paragraph}
% Examples of such defenses include CHERI~\cite{cheri}, Intel MPX~\cite{intelmpx}, Hardbound~\cite{hardbound}, WatchDog~\cite{watchdog}, WatchdogLite~\cite{watchdoglite}, CHEx86~\cite{chex86}, No-FAT (spatial safety)~\cite{no-fat},\mengjia{add spatial safety to figure 1} Intel CET~\cite{intelcet}, ARM BTI~\cite{armbti}, ZERO~\cite{zero}, REST~\cite{rest}, and Califorms~\cite{califorms}.\mengjia{can we remove califorms here to avoid confusion}

\paragraph{Taxonomy summary}
In this section, we have systematized a taxonomy of state-of-the-art memory corruption defenses focusing on their security against \ssb attacks. 
In this taxonomy, two categories of defenses (in red in \cref{fig:taxonomy}) may be vulnerable to \ssb attacks. 
Specifically, mitigations that augment tamperable metadata or add random offsets to addresses or pointers inevitably perform spoofable security checks; these spoofable security checks make these defenses sensitive to information disclosure attacks, since the leakage of secret metadata or offset empowers an adversary with the ability to disarm a security check and ultimately bypass the respective defenses.

On the other hand, the taxonomy has also led to finding of a class of defenses that are resilient to \ssb attacks; such defenses enforce the integrity of their metadata by techniques such as out-of-bound shadow memory, tagged memory, or page attributes. Furthermore, the metadata of these defenses are generally public information and indifferent to information disclosure.

\begin{figure*}[t]
    \centering
    \includegraphics[width=\textwidth]{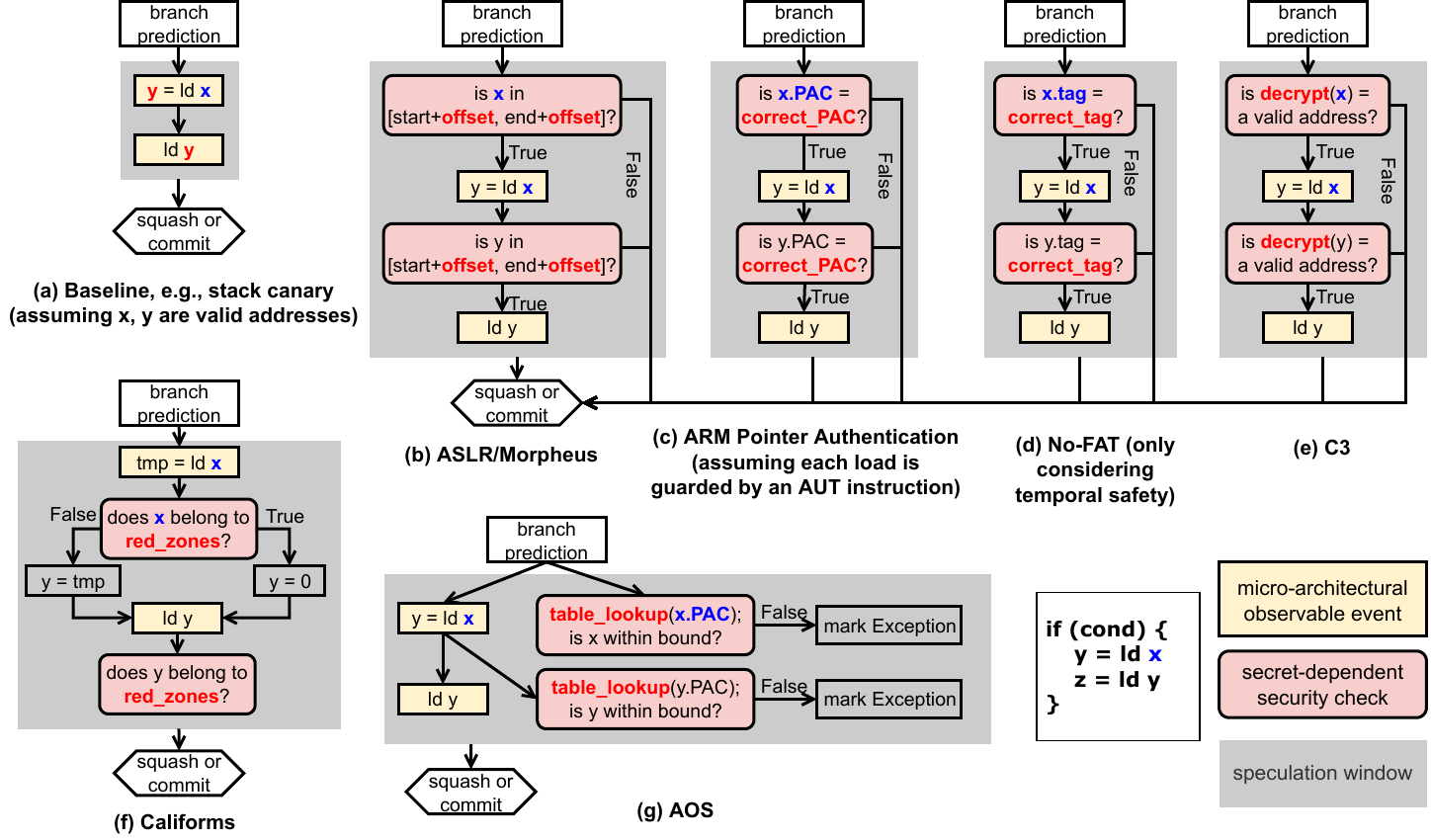}
    \caption{
    %Modeling speculative execution behaviors of memory corruption mitigations when executing the Spectre gadget. 
    Modeling speculative execution behaviors of memory corruption mitigations when executing a Spectre gadget.
    {\textcolor{red}{Red}} texts indicate secrets in the mitigation mechanisms, and {\textcolor{blue}{blue}} texts indicate attacker-controlled variables.}
    \label{fig:ssb_model}
\end{figure*}

\section{Modeling Speculative Shield Bypass Attacks}
\label{sec:model}

%We use a graph-based model to systematically analyze \ssb attacks on the 10 vulnerable memory corruption mitigations and potential future mitigations.
In \cref{sec:taxonomy}, we found two categories of defenses that may be vulnerable to \ssb attacks.
% , in the presence of an information disclosure attack that leaks some secret offset or metadata.
In this section, we use a graph-based model to systematically analyze the 10 defenses in these two categories.
%The goal is to capture the common characteristics of information leakage on different mitigations and understand how to defend against \ssb attacks.
The goal is to capture the common characteristics of information leakage of the different mitigations and ultimately better understand how to defend against \ssb attacks.

\subsection{Speculative Information Flow Graph}

%Recall that, in a speculative execution attack, the attacker triggers secret-dependent speculative execution in the victim program to result in distinguishable micro-architectural side effects.
Recall that in a speculative execution attack, an attacker triggers secret-dependent speculative execution in a victim program to result in distinguishable microarchitectural side effects.
%Therefore, our graph-based modeling approach focuses on tracking and visualizing secret-dependent information flows, including data-flow and control-flow dependencies at micro-architectural level.
Therefore, our graph-based modeling approach focuses on tracking and visualizing secret-dependent information flows, including data-flow and control-flow dependencies at the microarchitectural level.
%We call our graph a \graph (\shortgraph for short), because it literally captures information flow during speculative execution.
We call our graph a \graph (\shortgraph in short), because it captures the information flow during speculative execution.
% \mengjia{say it is simply information flow graph during speculative execution.}
%\joel{Maybe just "SIF graph" is better, since it seems to read a little more clearly to me.}

% \mengjia{it is unnecessary to invent completely new modeling method. we find our modeling renaissance... we show the key insights are almost the same, capturing execution ordering between operations, but we note the limitations of the original model only used for reasoning of leaking program data due to applying to different threat model in a discussion subsection.}

% Our graph shares a lot of similarities with the Topological Sort Graph used in \cite{ruby's understanding paper}.
In a \graph, a node represents a microarchitectural operation, which may or may not be observable via microarchitectural side channels.
% The granularity of the operations covered by each node is carefully chosen to capture the relationship that is critical for \ssb attacks without embedding unnecessary details. 
For example, a node could be a memory access operation, whose address can be distinguished via cache-based side channels, or a security check operation which is introduced by a mitigation mechanism.
%and could take various formats (\cref{sec:taxonomy}).
There exist two types of relationships between a pair of nodes.
First, if there exists a directed path from one node the the other, this means there exists a control-flow or data-flow dependency between the two operations, and the two operations are forced to happen sequentially in hardware.
Second, if there does not exist a directed path between the two nodes, this means there does not exist information flow between them and the two operations can happen in any order.
% Note that, different from traditional information flow graphs~\cite{xxx} that describe behaviors at program level from a software perspective, a \shortgraph describes relationships between micro-architectural operations from a hardware perspective.
% The two operations can not be exploited to leak information, since there does not exist information flow between them.

To clarify these relationships, we show the corresponding \shortgraph for each of the mitigation schemes in \cref{fig:ssb_model}.
The graphs describe the execution of the Spectre gadget \verb|if (cond) {y = ld x; z = ld y;}|.
We highlight the secrets in each mitigation scheme using red text, and the attacker-controlled variables using blue text.
The secrets in the mitigation mechanisms are either some tamperable metadata or the randomized offset.
The attacker-controlled variables include the variable $x$ and any metadata embedded inside it.
% We use these graphs to analyze whether and how the secrets can be leaked via \ssb attacks.

\subsection{Baseline Microarchitecture}
\label{sec:model:base}

%We start with analyzing the baseline micro-architecture to show that the \graph (\shortgraph) is simple and intuitive enough to help us understand how \ssb attacks work.
We start by analyzing the baseline microarchitecture. % to show that the \graph (\shortgraph) is simple and intuitive enough to help us understand how \ssb attacks work.
%We use yellow nodes to represent micro-architectural observable events and mark mis-speculation window using shades.
The mis-speculation window is shaded in gray and we use yellow nodes to represent microarchitectural observable events.
%Here, we assume programs are protected via software-only mitigations which store secret metadata in addressable virtual memory. 
We use stack smashing protection~\cite{stackguard_canaries} as an example. Stack smashing protection detects buffer overflows of local stack-based buffers, which overwrite the saved return address. 
By placing a random value (called a canary) between the return address and the local buffers at function entries, the integrity of the canary can be checked before the return of the function and thus the overflow can be detected.
Hence, the secret in the stack smashing protection is the canary.
%\mengjia{In S5, can we explain this?} \weon{added explanation here..}
% , which is stored in addressable virtual memory.

%In \cref{fig:ssb_model}(a), within the speculation window, we have two operations.
In \cref{fig:ssb_model}(a), within the speculation window, there are two operations.
The first operation performs a load using an attacker-controlled address $x$ and retrieves $y$.
%The value of $y$ can be the secret canary if the attacker makes the value of $x$ match the address of the location that stores the canary.
If the attacker controls the value of $x$ to equal to the address of the location that stores the canary, the value of $y$ would be the secret canary. 
%Next, the second load operation works as a transmission operation and leaks the value of $y$ (the canary) via cache side channels.
Next, the second load operation acts as a transmission operation and leaks the value of $y$ (the canary) via side channels.
Once the canary value is leaked, the attacker can spoof security checks and bypass stack smashing protection.
%We show a proof-of-concept demonstration of attacking the stack smashing protection on a real machine in \cref{sec:poc1}.
We demonstrate a proof-of-concept attack against stack smashing protection on a real machine in \cref{sec:poc1}.

\takeaway{\rev{If a mitigation is vulnerable to Spectre or its variants, and if the mitigation's metadata is stored in addressable virtual memory, then the secret metadata can be leaked via Spectre or its variants.}}

Memory corruption defenses introduce security checks to guard protected pointer de-reference operations.
% In our \shortgraph, we mark these security check operations as red nodes.
Interestingly, we found that defense mechanisms that leverage hardware-software co-design vary widely in their relationships between their security checks and their protected operations.
%We summarize three categories: 1) the security check sequentially guards the protected operation; 2) the security check modifies the behavior of the protected operation; 3) the security check happens in parallel with the protected operation.
We summarize these relationships into three categories: 1) the security check sequentially guards the protected operation; 2) the security check modifies the behavior of the protected operation; 3) the security check happens in parallel with the protected operation.

\subsection{Sequentially Guarded Security Checks}
\label{sec:model:seq}

We show the \shortgraph{s} for mitigations using sequentially guarded security checks in  \cref{fig:ssb_model}(b)-(e). %\pwd{This is a bit weird because sequentially guarded security checks is presented as though it's already been discussed in the taxonomy, but really isn't. Would it be worth adding it to Section 5?}
In these schemes, when executing the Spectre gadget, each of the two load operations is guarded with a security check node, meaning that the load operation will only be performed after the security check succeeds.
% if the security check fails, the load operation will not be performed.
%Taking ASLR (\cref{fig:ssb_model}(b)) as an example, the security check operation examines whether the address $x$ is mapped, that is, whether the address is within a valid region defined by the start and the end of the code or data segment and a randomized offset.
For example, in ASLR (\cref{fig:ssb_model}(b)), the security check operation examines whether the address $x$ is mapped; that is, whether the address is within a valid region defined by the start and the end of the code or data segment and a randomized offset.
%If $x$ is mapped, the load operation can be performed.
If $x$ is mapped, the load operation will be performed.
Otherwise, the load will be stalled and marked as illegal.
%The processor then waits for the branch to be resolved to decide either commit or squash these two instructions.
The processor then waits for the branch to be resolved to decide whether to commit or squash these two instructions.

We make the following observations from the \shortgraph{s}.
First, there exists an information flow from the security check node to the protected operation node, where the latter is an observable microarchitectural event. 
\rev{Hence, even if a mitigation's metadata resides in the virtual address space, it cannot be leaked via the attack vector described in \cref{sec:model:base}}.

\takeaway{\rev{If there exists an information flow between a preceding security check and a following observable microarchitectural event, then a memory corruption mitigation is able to mitigate Spectre and its variants.}}
% \jules{Here I am not sure I get it perfectly, even after several reads. 
% Are you talking about the specter gadget specifically here or about any type of observable microarchitectural event preceded by a security check?
% I am also confused by the use of "information flow" here and what observable microarchitectural event we are talking about (specter's transmitter?)
% If it's Specter specifically, maybe something like "if our security checks blocks further execution of the program, even speculatively, access to the transmitter in Specter-style attacks is made impossible and the memory-corruption mitigation effectively mitigates Specter and its variants"}

Second, the security check results in a boolean outcome where only one case results in an observable microarchitectural event.
Thus, the security check outcome can be leaked by observing whether the protected operation happens or not via microarchitectural side channels.
Note that, leaking the boolean security checkout outcome is sufficient for the attacker to spoof security checks.
This is because, with such a primitive, an attacker can abuse the security check to try all possible values of the attack-controlled variable (variable $x$) via brute-force, while suppressing exceptions via speculative execution, until the security check succeeds.
Once this value is leaked, the attacker can use this value to spoof the security checks.

%We can use the \shortgraph{s} to reason about existing attacks and easily derive attacks on unexploited mitigations.
\shortgraph{s} can be used to reason about existing attacks as well as derive new attacks.
%Those attacks that leverage speculative execution to bypass ALSR, such as RamBleed, speculative probing~\cite{speculive proging}, all follow the same pattern and exploit the control-flow dependency discussed above.
%The PACMAN attack~\cite{pacman} exploits the same pattern to bypass ARM Pointer Authentication, shown in \cref{fig:ssb_model}(c).
For example, Speculative Probing~\cite{speculativeprobing} and PACMAN~\cite{pacman} both exploit the information-flow dependency discussed above, to define appropriate ``transient-crash'' primitives, and bypass ASLR and ARM Pointer Authentication.
% No-FAT (temporal safety)~\cite{no-fat} and C3~\cite{c3} can be bypassed similarly.
We demonstrate a PoC attack targeting C3 in \cref{sec:poc3}, where we additionally identify a design flaw of C3 due to its use of a symmetric encryption algorithm.
To the best of our knowledge, we are the first to find vulnerabilities and demonstrate PoC attacks on C3.

\subsection{Califorms}
\label{sec:model:califorms}

%Califorms~\cite{califorms} is slightly different from the five mitigations discussed so far.
Califorms~\cite{califorms} differs to some degree from the five mitigations discussed so far.
We show the \shortgraph in \cref{fig:ssb_model}(f).
Instead of using security checks to guard protected operations, Califorms always speculatively execute the protected operations. 
\rev{However, to also mitigate Spectre, Califorms} uses the security check outcomes to influence the return values of the protected operations \rev{during speculation}.
Specifically, if the security check succeeds, the first load returns the correct data of $y$ and the second load will proceed as normal.
Otherwise, the value of $y$ will be set to $0$.

Despite the minor difference, Califorms shares an important common feature with the aforementioned schemes; that is, there exists an information flow from the security checks to the protected operations, and the speculative execution of protected operations can cause distinguishable microarchitectural side effects.
Therefore, a \ssb attack can leak the security check outcome in Califorms by monitoring the second load operation.
%To find the attack value that can be used to spoof the security check, the attacker will brute-forcely try all possible values for $x$ until the returned value $y$ becomes non-zero.
To find the attack value that can be used to spoof the security check, the attacker can try all possible values for $x$ by brute-force until the returned value $y$ is non-zero.

We summarize the attack vector discussed in \cref{sec:model:seq} and \cref{sec:model:califorms} below. 
% below.

\takeaway{As long as there exists an information flow from a security check to distinguishable microarchitectural events during speculative execution, an attacker can leak the security check  outcome via microarchitectural side channels. \rev{This phenomena arms an attacker with the ability to brute-force the attacker-controlled variable and find the secret value that can be used to spoof the security check. Note that, unlike the attack vector in \cref{sec:model:base}, this attack vector does not require the secret metadata/offset to be encoded inside the virtual address space.}
}

\paragraph{Entropy and Attack Bandwidth}
% In addition, the two attacks have very different attack bandwidth.
% Transient de-references directly leak the secret value, meaning multiple bits per leak.
% In comparison, the transient crashes 
The \ssb attack discussed above leaks a boolean value of the security check outcome, i.e. whether the check fails or not.
Thus, if a secret is n-bit, one must brute-force up to $2^n$ iterations.
%As the transient crash primitive requires the attacker to brute-force the secret, the entropy of these defense mechanisms largely determines the feasibility of such attacks.
As such a primitive requires the attacker to brute-force the secret, the entropy of these defense mechanisms largely determines the feasibility of such attacks.

Unfortunately, as AOS, ARM PA, No-FAT leverage unused upper bits of a pointer, the entropy of these mechanisms vary between 11
% \mengjia{I removed ARM ADI... so it should not start at 4...}\weon{changed to 11 for smallest PAC} 
and 16 bits, which is not enough to defend against brute-force attacks.
In Califorms,  the entropy of the size for each redzone is only 8 (i.e. 3 bits).
% \weon{Added back Califorms.}
The entropy of ASLR on a 64-bit system is between 19 bits (Windows) and 28 bits (Linux heap)~\cite{oregon-slides}.

%\mengjia{@Weon, put Morpheus here} \weon{added a draft}
Similar to ASLR, Morpheus randomizes the offset added to segments of the program. However, the entropy of Morpheus is 60 bits and the secret offset is re-randomized every 50ms. 
Thus, to brute-force the secret offset for Morpheus, one must be able to perform in-average $2^{59}$ attempts in under 50ms, which is quite challenging.
For example, according to PACMAN~\cite{pacman}, testing one boolean outcome takes around 2.69ms when the branch predictor is trained 64 times. 
As this implies that one can only try at most 18 brute-force attempts before re-randomization of the secret offset, we conclude that Morpheus is likely secure from brute-force attacks.
% \mengjia{how does the number work?}

\subsection{Parallel Security Checks}
\label{sec:model:parallel}

The third category of mitigations performs security checks in parallel with the protected operations during speculative execution.
We show the \shortgraph for AOS~\cite{aos} in \cref{fig:ssb_model}(g).
In AOS, the security check outcome does not interact with the speculative execution of the protected operations. 
Specifically, if the security check fails, the protected operation is still speculatively issued and only marked as causing an exception to be handled later.
%The exception is handled only when the instruction reaches the head of ROB, or will be ignored if the speculation window is squashed.
The exception is handled when the instruction reaches the head of ROB, but is ignored if the speculation window is squashed.

\takeaway{\rev{Parallel security checks lack information flow from a security check node to distinguishable microarchitectural events. Hence, they are resilient to the attack vector described in \cref{sec:model:seq} and \cref{sec:model:califorms}.}}

\rev{Despite the resiliency of AOS to the attack vector described in \cref{sec:model:seq} and \cref{sec:model:califorms},
%One might think that AOS is not vulnerable to \ssb attacks, since there is no information flow from any of the security check nodes to an observable microarchitectural event node.
%However, 
AOS is vulnerable to \ssb attacks. % due to a different vulnerability which we explain below.
To understand why, observe that} in \cref{fig:ssb_model}(g), the two load operations are always executed speculatively, independent from the outcomes of the security checks.
Hence, parallel security check based defenses are vulnerable to Spectre or its variants.
Since AOS is vulnerable to Spectre, arbitrary data in the virtual address space can be leaked speculatively. 
Furthermore, AOS persistently embeds the secret metadata, which is a hash (also called PAC), inside unused upper bits of the pointer.
%~\footnote{\rev{ARM PA and AOS use PACs differently. AOS generates a PAC for every heap pointer at allocation time which then persists until the allocation is freed; hence a heap pointer is stored with its PAC in virtual address space. ARM PA generates PACs for pointers at the beginning of critical sections of the program, but ROP/JOP gadget addresses generally do not include PACs because they were not initially signed; thus the secret PAC is generally not stored in the virtual address space.}}
We demonstrate a PoC attack targeting AOS in \cref{sec:poc2}.
To the best of our knowledge, we are the first to find vulnerabilities and demonstrate PoC attacks on AOS.

\takeaway{\rev{Parallel security checks lack information flow from a security check node to the following protected node. Hence, they are vulnerable to Spectre and its variants. Thus, if the metadata is encoded inside the virtual address space, they are vulnerable to the attack vector described in \cref{sec:model:base}.}}

%The lesson from AOS leads to the following takeaway message.

%\takeaway{
%To block \ssb attacks on a mitigation mechanism whose security checks can be spoofed, the two properties need to be \emph{both} satisfied: 1) do not store secret in addressable virtual memory, 2) disallow information flow from security checks to observable micro-architectural events.
% , and 3) disallow secret leakage inside the security check operation. 
%}

% \takeaway{
% To mitigate \ssb attacks, a defense which performs a spoofable security check must satisfy \emph{both} following properties: 1) not store secret in addressable virtual memory, and 2) disallow information flow from security checks to observable micro-architectural events.
% }

%\takeaway{Using parallel security checks has pros and cons. The benefit is that security check outcomes cannot be leaked speculatively. The downside is that the mitigation mechanism is vulnerable to Spectre and its variants.}

\subsection{Other Mitigation Mechanisms}

There are two mitigation mechanisms that are listed under spoofable security checks, but are not discussed above, that is, ARM MTE~\cite{armmte} and SPARC ADI~\cite{sparc_adi}.
According to public documentation, the secrets in these two schemes are memory tags, which are stored in specialized hardware structures.
However, there were no public documentation revealing enough microarchitectural details for us to derive their \shortgraph{s}.
%Considering that their secrets do not exist in addressable virtual memory, if they take the parallel security check implementation (similar to AOS~\cite{aos}) and their security check operation itself does not leak secrets, they will not be vulnerable to \ssb attacks.
In the case that they allow security check outcomes to affect observable microarchitectural events (similar to the first two categories of mitigations), they will be vulnerable to \ssb attacks.
On the other hand, if they use parallel security checks, considering that their secrets do not exist in addressable virtual memory, they will be secure from from \ssb attacks.
% these defenses may be secure from \ssb attacks if there exists no information flow from the security check nodes like AOS~\cite{aos}.\mengjia{could be clearer?}
%if they take the parallel security check implementation (similar to AOS~\cite{aos}) and their security check operation itself does not leak secrets, they will not be vulnerable to \ssb attacks.

%\subsection{Generality of the \shortgraph}
\subsection{Generality of \shortgraph{s}}
\label{sec:model:discuss}

% We discuss the generality of the \graph and how it differs from other graph-based representations for analyzing speculative execution attacks.

% \paragraph{Generality of the \graph}
%The \graph (\shortgraph) is an effective tool to reason about information flows between security checks and observable microarchitectural events.
\Graph{s} (\shortgraph{s}) are an effective tool to reason about information flows between security checks and observable microarchitectural events.
%So far, we have only used the graph to analyze these mitigation mechanisms using a Spectre gadget.
In this section, we have used \shortgraph{s} to analyze mitigation mechanisms using a Spectre gadget.
%It is completely feasible to use the graph to analyze other gadgets, such as using the pointers ($x$ and $y$) as branch targets for \texttt{jump} and \texttt{return} instructions.
It should be noted that \shortgraph{s} can also be used to analyze other varying gadgets, such as the ones which use the pointers ($x$ and $y$) as branch targets for \texttt{jump} or \texttt{return} instructions.

%Besides, even though we have only summarized three types of mitigations using \shortgraph{s}, the graph itself is general enough to capture different types of control-flow and data-flow dependencies. It is possible to use \shortgraph{s} to analyze new mitigations that could yield new information flow patterns.
Furthermore, even though we only summarize three categories of mitigations using \shortgraph{s}, the graph itself is, in fact, general enough to capture different types of control-flow and data-flow dependencies. 
%It is possible to use \shortgraph{s} to analyze new mitigations that could yield new information flow patterns.
In other words, \shortgraph{s} can be used to analyze new mitigations that  yield new information flow patterns yet to be discovered.

%Moreover, we used a single node to represent the security check operation for each mitigation mechanism in \cref{fig:ssb_model}, because these check operations are simple.
Moreover in \cref{fig:ssb_model}, we used a single node to represent the security check operation for each mitigation mechanism because these check operations are simple.
In the case that a security check operation becomes complex (e.g., requiring accessing complex data structures), we should consider decomposing a security check node into multiple nodes and investigate whether the security check itself introduces microarchitectural side effects that can leak a secret.
% secret leakage.

%Finally, we can use the \shortgraph to guide the design of countermeasures.
%Specifically, the existence of information flow paths leads to secret leakage (Takeaway 2), an effective countermeasure should aim to remove dependencies from secret-dependent security checks to observable microarchitectural events, which we discuss in \cref{sec:countermeasures}.
Finally, \shortgraph{s} can be used to guide the design of countermeasures,
% Specifically, as the existence of information flow paths leads to secret leakage (Takeaway 2), an effective countermeasure should aim to remove these dependencies from the secret-dependent security checks to observable micro-architectural events, 
which we discuss in \cref{sec:countermeasures}.

\section{Proof-of-Concept Demonstrations}
\label{sec:poc}

In this section, we demonstrate proof-of-concept (PoC) attacks against three memory corruption defenses that we have analyzed. 
%The three defenses are selected from different groups of mitigations in \cref{sec:vulnerabilityanalysis} to demonstrate diverse variations of synergistic attacks.
The first attack bypasses stack smashing protection~\cite{stackguard_canaries} using a Spectre primitive to leak the value of a \textit{stack canary}. 
% The second attack bypasses C3~\cite{c3} by leveraging a ``transient-crash'' primitive (\cref{sec:model:seq}) to brute-force a synonym cryptographic address of a target object on the heap. 
The second attack bypasses C3~\cite{c3} by leveraging the information flow between the security check nodes and the observable micro-architectural event nodes in its \shortgraph to brute-force a synonym cryptographic address of a target object on the heap. 
% \mengjia{I would like to remove this transient-crash term... since we only use it once}
Once we leak a synonym cryptographic address, we leverage Spectre to break data encryption.
The third attack bypasses AOS~\cite{aos} using a Spectre primitive to leak the PACs of objects on the heap.
%\mengjia{I reordered C3 and AOS. mention that we pick those that 1) the micro-architecture implementation is public (thus excluding MTE and ADI), 2) attack platform is avaialbe or open-sourced (excluding Califorms and NO-FAT), 3) Have not been attacked before (excluding ASLR, ARM PA).}
%weon{this is not true. (arguably) AOS and (definitely) C3 are also not public.}

In all the three PoC demonstrations, we set up a synthetic victim application, which contains 1) a memory corruption vulnerability, that allows either adjacent or non-adjacent buffer overflow; 
2) a Spectre gadget where the branch condition can be controlled by the attacker; 
3) a \texttt{win} function which resides in the victim's address space 
%but there exists no control-flow from the entry of the program to that 
lacking any control-flow entry from the program to that \texttt{win} function.
Our goal in the PoC demonstrations is to bypass each mitigation mechanism and demonstrate a successful control-flow hijack attack by overwriting a victim return address or a function pointer to the \texttt{win} function.

%Even though we target a simplified victim application, given that the three components are versatile in real-world applications, 
Despite the simplifications to our victim applications, given that these three components are common in real-world applications,
%our demonstration highlights that synergistic attacks are a real concern and future mitigations do need to take synergistic attacks into account.
we believe our PoC attacks promptly demonstrate the consequences of synergistic attacks in the wild.

\subsection{Breaking Stack Smashing Protection}
\label{sec:poc1}

%In our first PoC, we bypass stack smashing protection (SSP) using a spectre v1 primitive and an adjacent buffer overflow attack.
In our first PoC, we bypass stack smashing protection (explained in \cref{sec:model:base}) using a Spectre primitive and an adjacent buffer overflow vulnerability.
We conduct our experiment on an Intel Xeon Gold 5220R CPU @ 2.20GHz, running Ubuntu 18.04.6 LTS.
%In a SSP-protected program, at the function prologue, a random value, called a canary, is placed onto the stack adjacent to the saved return address.
%At the function epilogue, before the return address is used, the integrity of the saved canary is validated in software by checking whether the canary is the same.
%In our 64-bit system, a canary is 8 bytes. 
%The same canary is used throughout the lifetime of the program.
In our PoC, we undertake two steps.

\paragraph{Step 1: Using a Spectre primitive to leak the canary}
%We follow the standard Sepctre v1 attack (gadget below) to leak the canary value byte-by-byte.
According to the \shortgraph in \cref{fig:ssb_model}(a), the canary can be leaked via Spectre attacks.
Thus, we leverage a standard Spectre primitive to leak the canary byte-by-byte using a Flush+Reload attack.
% We first train the branch predictor to be taken, and then trigger a speculative out-of-bound access to make \texttt{array1[x]} point to the canary on the stack.
% We then conduct a Flush+Reload attack to detect which element in \texttt{array2} was accessed and infer the value of the corresponding canary byte. 
We repeat the attack 8 times to leak all 8 bytes of the canary, which can be used to spoof the security check.
% \begin{minted}[frame=single,fontsize=\footnotesize,      obeytabs=true,tabsize=4,numbersep=3pt,linenos,escapeinside=||]{c}
% if (x < array1_length): // x is attacker-controlled
%   uint8_t y = array1[x]    // array1[x] points to canary
%   uint8_t z = array2[y*64] // transmit one byte of canary
% \end{minted}

%Step 2: overwrite the return address on the stack to the address of the \texttt{win} function.
\paragraph{Step 2: Overwriting the return address on the stack to the address of the \texttt{win} function}
%Since we exploit an adjacent buffer overflow vulnerability (the \texttt{gets} example in \cref{sec:bac_memattack}), before overwriting the return address, we have to first overwrite the canary.
The limitations of an adjacent buffer overflow vulnerability (the \texttt{gets} example in \cref{lst:mem-corrupt}) requires us to overwrite the canary, before overwriting the return address.
%We embed the correct canary value leaked from step 1 in our payload, so that the canary value is not changed after being overwritten, and thus SSP is bypassed.
%The program then executes a \texttt{ret} instruction that uses the modified return address and triggers the \texttt{win} function, declaring that our control-flow hijack attack succeeds.
Thus, we embed the correct canary value that we leaked during step 1 into our payload, so that the canary is the same value after being overwritten, while the return address is modified to point to \texttt{win}. %and SSP is bypassed.
At the function epilogue, the modified (but same value) canary passes validation and the program then executes a \texttt{ret} instruction that uses the modified return address. This triggers the \texttt{win} function, declaring our control-flow hijack attack to be successful.

\subsection{Breaking C3}
\label{sec:poc3}

In our second PoC, we bypass C3~\cite{c3}.
%. leveraging a combination of transient-crash and Spectre v1 primitives.
%mixture of transient-crash primitive and the spectre v1 primitive.
C3 is built using two key techniques, \emph{address encryption} and \emph{data encryption}, which we break in steps.
To break address encryption, we abuse the control-flow dependence between the security check and protected memory operation (\cref{sec:model:seq}). 
% \mengjia{no need to keep this term.}
To break data encryption, we additionally identify another design flaw of C3, namely symmetric encryption of data.
Lacking a real-world C3 implementation, we implement C3 using an augmented allocator and an architectural simulator.
We evaluate our attack in gem5~\cite{gem5}.
%in the memory allocator \pwd{this is the first reference to this memory allocator, might want to elaborate} and in the gem5 simulator~\cite{gem5}.
%\mengjia{mention we find a design flaw ....}

\begin{figure}[h]
    \centering
    \includegraphics[width=.85\columnwidth]{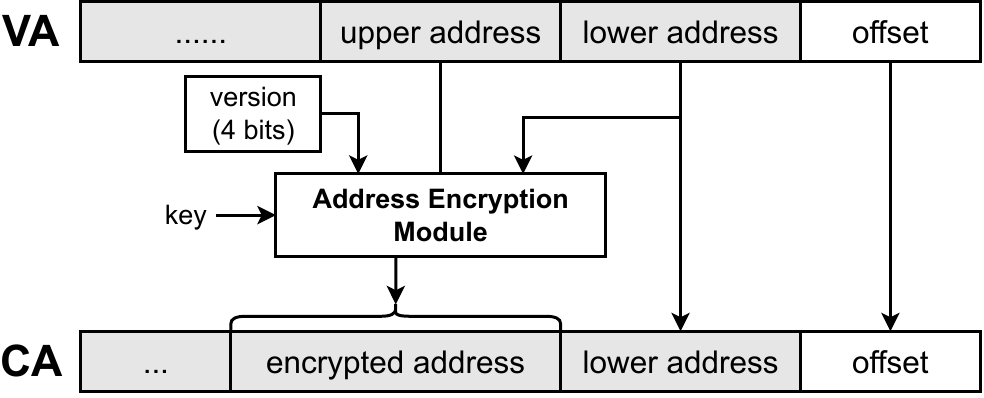}
    \caption{Address Encryption in C3.}
    \label{fig:c3-address-encryption}
\end{figure}

\paragraph{Address Encryption}
%For each heap allocation, C3 uses the obtained virtual address (VA) to generate a cryptographic address (CA for short), which can be used as a unique signature for that allocation.
For each heap allocation, C3 obtains a virtual address (VA) to generate a cryptographic address (CA for short). 
CAs act as a unique signature for that allocation.
\cref{fig:c3-address-encryption} shows the CA computation process.
A virtual address (VA) is divided into three parts: the offset, the lower address, and the upper address.
The lower address and the upper address, as well as a 4-bit version number and a secret key are used as inputs to a block cipher (K-cipher~\cite{kcipher}) to generate an encrypted address.~\footnote{We omit several technical details regarding C3 that are less relevant to our proof-of-concept, such as the power.}
%the encrypted address.~\footnote{We omitted several technical details about C3 that are not relevant to our attack, such as the power value.}
The decryption process is reversed.

%The key idea of C3 is that if an attacker tampers any bits other than the offset, the tampered CA will likely be decrypted as a garbled address and de-referencing a garbled address will cause an exception.
% The key insight behind the address encryption of C3 is that if an attacker tampers with any bits other than the offset, the tampered CA will likely be decrypted as a garbled virtual address that causes an exception and leads the program to crash. 
%and de-referencing a garbled address will cause an exception.
% Hence, our transient crash attack vector, which takes advantage of speculative exceptions to leak the outcome of security checks, can be leveraged to break the address encryption of C3.
%Recall from \cref{sec:attackvector}, our transient crash attack vector takes advantage of speculative exceptions to leak the outcome of security checks.

%that primitive can be used to detect and is a perfect fit to break C3's address encryption.

\paragraph{Breaking Address Encryption}
According to the \shortgraph in \cref{fig:ssb_model}(e), an attacker can exploit the information flow between the security check and protected memory operations to leak the boolean security check outcome.
Specifically, in our PoC, we aim to use the security check outcome to distinguish between CAs that decrypt to a specified VA and CAs that result in garbled VAs, which we call a \emph{cryptographic address oracle}.
% In our PoC, we construct a \emph{cryptographic address oracle} to distinguish between CAs that decrypt to a specified VA and CAs that result in garbled VAs.
With such an oracle, an attacker is able to brute-force all possible CAs that decrypt to a specified VA.
%they find the correct ones.
Note that, due to the 4-bit version number, there exist 16 different CAs that map to the same VA after decryption.

%In our PoC, we construct our cryptographic address oracle using a gadget snippet like the one below.
We use the gadget snippet like the one below to construct our cryptographic address oracle.
Specifically, by transiently brute-forcing \texttt{CA\_guess}, the attacker tries to forge an address called \texttt{CA\_attack} that decrypts to the same VA as the victim's address \texttt{CA\_victim}.
We refer to \texttt{CA\_attack} as a \emph{synonym CA} of \texttt{CA\_victim}.
\begin{minted}[frame=single,fontsize=\small,      obeytabs=true,tabsize=4,numbersep=3pt,linenos,escapeinside=||]{c++}
struct victim { void* fp = &(correct_func()); }
CA_victim = new victim;
if (cond) {  dummy = *CA_guess; }
\end{minted}
%\pwd{The struct isn't defined in Listing 2?}
%weon: thanks!!!

We first train the branch to be taken, and trigger a speculative de-reference of \texttt{CA\_guess}.
We then use Prime+Probe to monitor the cache set that virtual address of \texttt{CA\_victim} maps to and count the number of cache misses.
%\mengjia{virtual address CA\_victim???}
In our experiment, we are able to precisely distinguish incorrect CAs from synonym CAs.
Specifically, when the guess is a synonym CA, our Prime+Probe implementation in gem5 observes 3 cache misses.
Otherwise, 
% instead of performing a cache access, the speculative de-reference results in a speculative crash, and 
our implementation observes 2 cache misses.
%~\footnote{Our implementation distinguishes between 2 versus 3 cache misses, not 0 versus 1 miss. We suspect that this happens due to the default cache replacement policy in gem5.}

\paragraph{Data Encryption}
In addition to address encryption, C3 uses data encryption to further enhance its protection coverage.
The data encryption module performs an XOR operation that can be summarized as below, where \texttt{PRP} is a keystream generated using a pseudo-random permutation of the cryptographic address (CA).
The decryption process is reversed.
With such an encryption scheme, a CA becomes a unique key for data of each allocation.

\vspace{-1ex}
\begin{center}
    \texttt{Data\_plaintext $\oplus$ PRP(CA) => Data\_encrypted}
\end{center}
\vspace{-1ex}

%The key idea of C3 data encryption is that even if the attackers can find a synonym address to access a VA, they will still get garbled data.
% The key insight behind data encryption of C3 is that even if attackers can forge a synonym CA that decrypts to a specific VA, the data will still be garbled.
%they will still get garbled data.
For example, assume \texttt{CA1} and \texttt{CA2} are synonym CAs that decrypt to the same VA. If plaintext data is encrypted using CA1, but then decrypted using CA2, the result is garbled data as shown below.
%it will result in garbled data as shown below.

\vspace{-1ex}
\begin{center}
\texttt{
Data\_plaintext $\oplus$ PRP(CA1) => Data\_encrypted\\
Data\_encrypted $\oplus$ PRP(CA2) => Data\_garbled \qquad}
\end{center}
\vspace{-1ex}

\paragraph{Breaking Data Encryption}
To bypass data encryption, we exploit the weakness of the symmetric encryption via XORs.~\footnote{We note that our attack is specific to the weakness of XOR encryption.
If C3 were to upgrade to a more advanced encryption algorithm, it may require us to acquire the same CA, not just a synonym CA, during our attack on address encryption,
%get the exact same address when breaking the address encryption in the first step
increasing the attack difficulty. 
% \pwd{Just 16x right? Is that a signficant difference?}\mengjia{it is ok to remove significantly}
}
Our insight is that according to the algorithm above, if the attackers are able to leak one pair of plaintext data and the corresponding garbled data, they can derive a mask (like the snippet below) that can be used to directly translate between any possible pairs of plaintext data and garbled data without 
%knowing anything about the random permutation keystream.
any knowledge regarding the keystream.
%the relationship between any pair of plaintext data and garbled data, as below.

\vspace{-1ex}
\begin{flushleft}
\qquad\texttt{mask = Data\_plaintext $\oplus$ Data\_garbled \\
\qquad\qquad\  = PRP(CA1) $\oplus$ PRP(CA2)}
\end{flushleft}
\vspace{-1ex}

\pwd{It might be useful to fully flush out the example by demonstrating how the mask works with an equation}
Our PoC attack works as follows, shown in \cref{lst:c3-attack}.
%Similar as before, the victim allocates a struct on heap and obtains \texttt{CA\_victim}, which points to a function pointer \texttt{fp} that we try to overwrite (line 1).
The victim allocates a struct \texttt{victim} (similar to \cref{lst:victim}) on heap and obtains \texttt{CA\_victim}, which points to a function pointer \texttt{fp} that we will try to overwrite (line 1).
First, following the approach described above to break address encryption, we obtain \texttt{CA\_attack} which is a synonym CA of \texttt{CA\_victim}.
Second, we use a Spectre primitive to leak the garbled data inside the function pointer \texttt{fp} (lines 2-5).
% , byte-by-byte by speculatively de-referencing \texttt{CA\_attack} (lines 2-5).
Third, after we leak the garbled data, which we denote as \texttt{garbled\_fp}, we compute the mask that can be used to translate between garbled data and plaintext data.
Next, the \texttt{mask} is XORed with the address of the \texttt{win} function to derive the intermediate data, and we overwrite \texttt{fp} using this intermediate data and \texttt{CA\_attack} (line 8).
Finally, as the \texttt{fp} is successfully overwritten, when decrypted during function call using \texttt{CA\_victim}, the control flow is successfully directed to our \texttt{win} function (line 9).

\begin{listing}

\begin{minted}[frame=single,fontsize=\small,      obeytabs=true,tabsize=4,numbersep=3pt,linenos,escapeinside=||]{c++}
CA_victim = new victim; // CA_victim points to fp
if (cond) { // Spectre gadget
    garbled_byte = *CA_attack;
    dummy = array[garbled_byte * 64];
}
// compute the data to overwrite fp
mask = fp_garbled ^ fp;
*CA_attack = addr_win ^ mask;
call CA_victim.fp(); // call-to-win
\end{minted}
\vspace{-4ex}
\caption{Attacking data encryption of C3.}
\label{lst:c3-attack}
\end{listing}

\subsection{Breaking AOS}
\label{sec:poc2}
%In our second PoC, we bypass AOS using a spectre v1 primitive and a non-adjacent buffer overflow attack.
In our third PoC, we bypass AOS using a Spectre primitive and a non-adjacent buffer overflow vulnerability.
We conduct our experiment using an architectural simulator and an augmented compiler. %the gem5 simulator~\cite{gem5}. 
Specifically, the hardware implementations~\cite{aos-gem5} are in gem5~\cite{gem5} and the compiler implementations~\cite{aos-llvm} are in LLVM~\cite{llvm}. 

\paragraph{AOS for Heap Memory Safety}
% Always On Heap Memory Safety (AOS) \pwd{You spell out the acronym here for the first time, probably should do that earlier} 
% AOS is a memory safety solution designed to provide temporal and spatial memory safety for heap allocations.
In AOS, for every heap allocation, the compiler generates a pointer authentication code (PAC), which is a cryptographic hash
%computed in the same way as ARM PA~\cite{armpa}
, and stores the PAC in the upper unused bits of the pointer.
Additionally, the bound information of the pointer is stored in a Hashed Bounds Table (HBT) indexed using the generated PAC.
Upon de-referencing a pointer, the AOS hardware extracts the PAC from the pointer,  uses the PAC to index into the HBT to retrieve the bound information, and performs a bound-check operation.

\begin{figure}[h]
    \centering
    \includegraphics[width=\columnwidth]{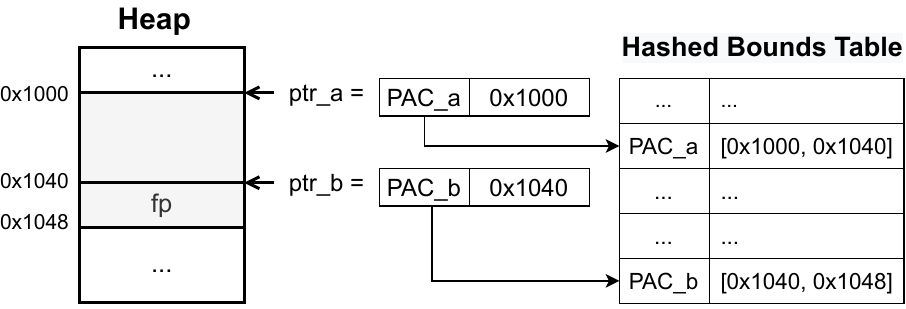}
    \caption{AOS protection example.}
    \label{fig:aos_example}
\end{figure}

To understand how AOS works, consider the following example in \cref{lst:victim} and the corresponding heap content and HBT status in \cref{fig:aos_example}.
In this example, the \texttt{victim} struct contains a function pointer that the attacker wants to overwrite.
The victim program allocates an array at address \texttt{0x1000} and a \texttt{victim} struct at address \texttt{0x1040} on the heap.
The obtained \texttt{ptr\_a} and \texttt{ptr\_b} will have PAC values embedded in the high bits, denoted as \texttt{PAC\_a} and \texttt{PAC\_b}, respectively.
When the attacker uses the vulnerability on line 4 to perform an out-of-bound access (i.e., setting \texttt{x} larger than array size), the AOS hardware will capture it, because the PAC value of \texttt{ptr\_a[x]} will be used to do a HBT lookup and a bound check will detect such illegal bound-bypass activity.

\vspace{-5pt}
\begin{listing}[h]
\begin{minted}[frame=single,fontsize=\small,      obeytabs=true,tabsize=4,numbersep=3pt,linenos,escapeinside=||]{c++}
struct victim { void* fp = &(correct_func()); }
ptr_a = new uint64_t[8]; // bound: 0x1000 to 0x1040
ptr_b = new victim;      // bound: 0x1040 to 0x1048
ptr_a[x] = y; //x, y are 64-bit, attacker-controlled
call ptr_b.fp();
\end{minted}
\vspace{-4ex}
\caption{Victim code snippet.}
\vspace{-1ex}
\label{lst:victim}
\end{listing}\vspace{-6pt}

\vspace{-6pt}

\paragraph{AOS Vulnerability}
As discussed in \cref{sec:taxonomy}, AOS uses tamperable metadata and its security check can be spoofed.
% the PAC bits stored with the pointer are tamperable.
Attackers can bypass AOS's protection if they can modify a pointer together with its PAC to match another valid pointer and its associated PAC.
Using the same example from \cref{lst:victim}, an out-of-bound write on line 4 could succeed if the index \texttt{x} can set the PAC bits of \texttt{ptr\_a} to match \texttt{PAC\_b}.
As a result, the bound information will be retrieved from the HBT using \texttt{PAC\_b} as index, and the bound check will go through.
To do so, the attacker needs to first leak the PAC values.

\paragraph{Attacking AOS}
In our third PoC, we set up our synthetic victim program similar to the code in \cref{lst:victim}.
%We take two steps to break AOS.
We undertake two steps to break AOS.

\paragraph{Step 1: Using a Spectre primitive to leak PAC values}
According to the \shortgraph in \cref{fig:ssb_model}(g) and the discussion in \cref{sec:model:parallel}, AOS is vulnerable to Spectre and its variants due to parallel security checks which fail to protect load operations during speculative execution. 
% Specifically, during speculation, AOS performs HBT lookups in parallel with memory accesses to achieve better performance. Thus, upon HBT bound-check
% failure, the corresponding instruction is marked as a faulting
% instruction, but the exception is only handled when the triggering instruction reaches the head of ROB. 
Our experiments show that a Spectre primitive works effectively on an architectural simulator of AOS, and that a Prime+Probe attack on the cache can  precisely leak the two PACs, \texttt{PAC\_a} and \texttt{PAC\_b}.

\paragraph{Step 2: Overwriting a function pointer on the heap to the address of the \texttt{win} function}
%We aim to modify the \texttt{fp} value inside the \texttt{victim} struct by exploiting the non-adjacent buffer overflow vulnerability on line 7 in \cref{lst:victim}.
We then modify the \texttt{fp} inside the \texttt{victim} struct by exploiting the non-adjacent buffer overflow vulnerability on line 4 in \cref{lst:victim}.
To bypass AOS protection, we calculate the index \texttt{x} by subtracting \texttt{PAC\_a} from  \texttt{PAC\_b} and then add the offset to make the resulted pointer points to the \texttt{victim} struct
%~\footnote{In the case that \texttt{x} has a limited bitwidth and cannot affect the PAC bits, our attack can still work via leveraging memory spraying~\cite{wiki-heap-spraying} and exploiting PAC collision~\cite{aos}.}.
%\mengjia{if you subtract between ptrs, you do not need to add the offset..}
As the PAC of \texttt{ptr\_a[x]} becomes \texttt{PAC\_b}, the bound check is spoofed to pass, and \texttt{ptr\_b.fp} is modified to the address of the \texttt{win} function.
Later when the tampered function pointer is called, the \texttt{win} function is successfully triggered,
declaring our control-flow hijack attack to be successful.

\section{Countermeasures}
\label{sec:countermeasures}

We present potential countermeasures, derived from the insights in \cref{sec:taxonomy} and \cref{sec:model}.

First, as discussed in \cref{sec:taxonomy},  tamperproof metadata augmentation based schemes are not vulnerable to \ssb attacks as their security checks cannot be spoofed. As such, one can incorporate techniques used by these mechanisms to enforce metadata integrity.
% Proposed sentence to say this is not perfect neither:
%Though those schemes also have limitations, they are often invasive both to hardware and software and have seen few deployments.

\rev{

Second, one can leverage the SIF graphs from \cref{sec:model} to
reason about possible countermeasures for spoofable security checks.
For example, for the schemes that are insecure due to the leakage of security check outcomes (\cref{fig:ssb_model}(b)-(f)), one can try to remove the edges from the graph, that is, to remove the information flow paths from security check nodes to observable microarchitectural event nodes.
In other words, one could alter the \shortgraph{s} to perform parallel checks, similar to AOS (\cref{fig:ssb_model}(g)).
This would prevent an attacker from leaking the security check outcome for \cref{fig:ssb_model}(b-f).
%In fact, shortly after the publication of PACMAN~\cite{pacman}, ARM has recommended such a patch for ARM PA~\cite{arm-pacman-article}.
In fact, shortly after the publication of PACMAN~\cite{pacman}, ARM suggested such a patch~\cite{arm-pacman-article}.
However, such an approach may not be straightforward for many defense mechanisms.
For instance, for ASLR, Morpheus, and C3, if the security check fails, this would mean that the address is unmapped and it is unclear what data should be returned. In this case, one could consider returning data from another mapped address instead.

For example, to protect code pointers against SSBs in ASLR, one can compare each instruction address fetched with the base and bound of the code segment of the program.
If the address is within the base and bound of code segment, the address is mapped and the original address can be fetched.
If the address is not, one can instead fetch the modulo of the requested address by the code segment size, added to the base.
When no attack is happening, this scheme only adds one comparison.
As such, this scheme would incur at most one extra cycle of latency when fetching an instruction.
However, such an approach would lead to Spectre vulnerabilities (\cref{sec:model:parallel}) and thus any secrets related to the ASLR scheme should not be stored in the virtual address space.
To achieve this, one can either generate the metadata only during critical sections of the program~\cite{qualcomm_white_paper} or leverage specialized hardware to maintain the metadata~\cite{no-fat, armmte, sparc_adi}.

A more sophisticated approach would be to explore integrating prior works that mitigate information leakage from speculative microarchitectural events.
Mitigation mechanisms that hide cache modulations caused by speculative memory accesses, such as \cite{invisispec,safespec,dom,muontrap} could serve this purpose. In terms of SIF graphs, such direction would be equivalent to transforming all
micro-architectural events (the yellow nodes in \cref{fig:ssb_model}) into unobservable events (grey nodes).
In other words, such a transformation would not only block Spectre from leaking secrets that are stored in virtual memory (\cref{fig:ssb_model}(a), (g)), but also block leaking security check outcomes (\cref{fig:ssb_model}(b)-(f)).
STT~\cite{yu2019stt} would also mitigate leakage of metadata from virtual memory (\cref{fig:ssb_model}(a), (g)), but cannot block leakage of security check outcomes (\cref{fig:ssb_model}(b)-(f)) as it allows the first load to propagate during speculation. However, we expect integrating such mechanisms with memory corruption defenses to be non-trivial.
%Proposed sentence
% One could consider that the architects proposing speculative mitigations should be expected to think about their new speculative blocking scheme not only against standard speculative threat model of a program that does not have vulnerabilities, but against an SSB threat model. 
% Different speculative mitigative might yield different security guarantees under this new threat model.

}

\rev{
\section{Discussion: Threat Models}
}
\label{sec:discussion}

\rev{
Traditional speculative-execution attacks, memory corruption attacks and \ssb attacks differ in the underlying assumptions of their respective threat models, and in the attacker's end goals.
Traditional speculative-execution attacks assume secure software, but hardware that is vulnerable to side-channels. Hence, as noted in \cref{sec:bac_side_channel}, in a traditional speculative-execution attack, the attacker ultimately aims to violate confidentiality by leaking program data in the victim program's security domain.
Conversely, memory corruption attacks assume secure hardware, but buggy software. 
Thus, as noted in \cref{sec:bac_memattack}, memory corruption attacks, in addition to leaking program data, aim to violate \emph{integrity} by modifying the program data, ultimately compromising authority and performing arbitrary code execution.

\Ssb attacks assume both side-channel-vulnerable hardware and buggy software.
Under this assumption, \ssb attacks not only aim violate confidentiality by leaking program data, but in-line with memory corruption attacks, aim to violate integrity and modify program data, ultimately to gain arbitrary code execution with higher authority.
Thus, under a speculative-shield-bypass threat model, the attacker's end goal is to perform a successful memory corruption attack. However the presence of a memory corruption mitigation mechanism necessitates an intermediate step; that is first gaining the ability to bypass the defense.

In this context, the primary question for a \ssb attack is whether whatever security check imposed by a defense is spoofable.
If so, the speculative component of a \ssb attack does not aim to read arbitrary program data, but rather uniquely targets the defense metadata to facilitate the more powerful memory corruption attack. Note that while some defense schemes maintain their metadata inside the virtual address space~\cite{stackguard_canaries, pax_aslr, morpheus, aos}, making the distinction between program data and metadata obscure, in many defense schemes this is not true~\cite{armpa, no-fat, c3, armmte, sparc_adi}.

There is a complex interaction between the properties of a memory corruption defense and its vulnerabilities to varying threat models.
As noted in \cref{sec:model:seq}, while most memory defenses were not designed to protect against traditional speculative-execution attacks, the extra security checks sometimes break the standard transmitters used by speculative-execution attacks.
For example, in \cref{fig:ssb_model}(b)-(e), spectre transmitters are broken because any load would be blocked by the defense scheme's during speculation.

In the same section, we also showed that while the security checks imposed by the memory-corruption defenses in \cref{fig:ssb_model}(b)-(e) block standard transmitters, they introduce new limited transmitters.
These cannot transmit arbitrary program data, but they can transmit the metadata of memory-corruption defenses, which lead to \shortssb vulnerabilities.

Reciprocally, some memory-corruption mitigations, while not vulnerable to \shortssb because their metadata are tamperproof, are unable to block standard speculative-execution attacks. 
For instance, consider CHERI~\cite{cheri}. Because the integrity of its metadata is enforced, CHERI's security checks are unspoofable, and hence CHERI is resilient to \ssb attacks. 
However, CHERI performs parallel security checks similar to AOS in \cref{fig:ssb_model}(g). 
Thus, as explained in \cref{sec:model:parallel}, CHERI is unable to protect its program data from Spectre and its variants. 
In fact, most tamperproof metadata defenses opt to perform parallel security checks schemes (for performance reasons), making them vulnerable to Spectre and its variants, including \cite{intelmpx,hardbound,watchdog, watchdoglite,chex86,rest} and~\cite{califorms}.

We summarize security trade-offs of memory-corruption-mitigation schemes between the traditional-speculative-execution threat model and speculative-shield-bypass threat model in \cref{tab:spectre}. 
Column 3 and 4 indicate whether program data or the metadata/offset can be leaked via traditional speculative-execution attacks. Column 5 indicates whether an attacker with a memory write vulnerability is able to corrupt the metadata/offset. Column 6 indicates whether the metadata/offset is brute-forceable by abusing security checks. Column 7 indicates whether the respective memory corruptions are vulnerable to \shortssb attacks. This requires the metadata to be corruptable and, if so, either vulnerable to traditional speculative-execution attacks or brute-forceable.

%\begin{figure*}[t]
%    \centering
%    \includegraphics[width=0.7\textwidth]{figures/table-spectre.png}
%    \caption{Spectre vs SSB}
%    \label{tab:spectre}
%\end{figure*}

% \usepackage{multirow}
\begin{table*}[t]
\footnotesize
\begin{center}
\begin{tabular}{|c|c|cc|cc|c|}
\hline
\multirow{2}{*}{Classification} &
  \multirow{2}{*}{Proposal} &
  \multicolumn{2}{c|}{Vulnerable to Spectre+} &
  \multicolumn{2}{c|}{Metadata/Offset} &
  \multirow{2}{*}{\begin{tabular}[c]{@{}c@{}}Vulnerable\\  to SSB\end{tabular}} \\ \cline{3-6}
 &
   &
  \multicolumn{1}{c|}{Program Data} &
  Metadata/Offset &
  \multicolumn{1}{c|}{Corruptable} &
  Brute-forceable &
   \\\hline
\multirow{3}{*}{\begin{tabular}[c]{@{}c@{}}Tamperproof\\ Metadata\end{tabular}} 
% &
%   ARM BTI/Intel CET &
%   \multicolumn{1}{c|}{\checkmark} &
%   Yes, but public &
%   \multicolumn{1}{c|}{x} &
%   x &
%   x \\
 % &
 %  Intel CET &
 %  \multicolumn{1}{c|}{\checkmark} &
 %  Yes, but public &
 %  \multicolumn{1}{c|}{x} &
 %  x &
 %  x \\
 
 &
  \begin{tabular}[c]{@{}c@{}}CHERI/Intel MPX/Hardbound\\Watchdog/Watchdoglite/ CHEx86\\ ARM BTI/Intel CET/ REST\end{tabular}
   &
  \multicolumn{1}{c|}{\checkmark} &
  Yes, but public &
  \multicolumn{1}{c|}{x} &
  x &
  x \\
  &
  ZERO/No-FAT (spatial) &
  \multicolumn{1}{c|}{x} &
  x &
  \multicolumn{1}{c|}{x} &
  x &
  x \\
  % &
  % \begin{tabular}[c]{@{}c@{}}No-FAT (spatial)\end{tabular} &
  % \multicolumn{1}{c|}{x} &
  % x &
  % \multicolumn{1}{c|}{x} &
  % x &
  % x \\
  \hline
\multirow{2}{*}{\begin{tabular}[c]{@{}c@{}}Address Layout\\ Randomization\end{tabular}} &
  ASLR/Califorms (non-adjacent) &
  \multicolumn{1}{c|}{x} &
   x &
  \multicolumn{1}{c|}{\checkmark} &
  \checkmark &
  \checkmark \\
 &
  Morpheus &
  \multicolumn{1}{c|}{x} &
   x &
  \multicolumn{1}{c|}{\checkmark} &
  x &
  x \\
 % &
 %  \begin{tabular}[c]{@{}c@{}}Califorms (non-adjacent)\end{tabular} &
 %  \multicolumn{1}{c|}{x} &
 %   x &
 %  \multicolumn{1}{c|}{\checkmark} &
 %  \checkmark &
 %  \checkmark \\ 
  \hline
\multirow{3}{*}{\begin{tabular}[c]{@{}c@{}}Tamperable \\ Metadata\end{tabular}} &
  Stack Canary/AOS/No-FAT (temporal) &
  \multicolumn{1}{c|}{\checkmark} &
  \checkmark &
  \multicolumn{1}{c|}{\checkmark} &
  x &
  \checkmark \\
  % &
  % AOS &
  % \multicolumn{1}{c|}{\checkmark} &
  % \checkmark &
  % \multicolumn{1}{c|}{\checkmark} &
  % x &
  % \checkmark \\
  %  &
  % \begin{tabular}[c]{@{}c@{}}No-FAT\\ (temporal)\end{tabular} &
  % \multicolumn{1}{c|}{\checkmark} &
  % \checkmark &
  % \multicolumn{1}{c|}{\checkmark} &
  % x &
  % \checkmark \\ 
 
 &
  ARM MTE/ SPARC ADI &
  \multicolumn{1}{c|}{?} &
  ? &
  \multicolumn{1}{c|}{\checkmark} &
  ? &
  ? \\
 % &
 %  SPARC ADI &
 %  \multicolumn{1}{c|}{?} &
 %  ? &
 %  \multicolumn{1}{c|}{\checkmark} &
 %  ? &
 %  ? \\
 
 &
  C3/ ARM PA &
  \multicolumn{1}{c|}{x} &
  x&
  \multicolumn{1}{c|}{\checkmark} &
  \checkmark &
  \checkmark \\
  % &
  % ARM PA &
  % \multicolumn{1}{c|}{x} &
  % x &
  % \multicolumn{1}{c|}{\checkmark} &
  % \checkmark &
  % \checkmark \\
\hline
\end{tabular}
\caption{\rev{Memory corruption mitigations under traditional-speculative-execution and speculative-shield-bypass threat models}}
\label{tab:spectre}
\end{center}
\end{table*}

%====================
%
%summarize table.
%
%===================

%\begin{figure*}[t]
%    \centering
%    \includegraphics[width=0.8\textwidth]{figures/table-safety.png}
%    \caption{Memory Safety Error Detection}
%    \label{tab:safety}
%\end{figure*}

% \usepackage{multirow}
% \usepackage{multirow}

% \usepackage{multirow}
\begin{table*}[t]
\footnotesize
\begin{center}
\begin{tabular}{|c|c|cc|c|c|c|c|}
\hline
\multirow{2}{*}{Classification} &
  \multirow{2}{*}{Proposal} &
  \multicolumn{2}{c|}{Spatial Protection} &
  \multirow{2}{*}{\begin{tabular}[c]{@{}c@{}}Temporal\\ Protection\end{tabular}} &
  \multirow{2}{*}{Maturity} &
  \multirow{2}{*}{\begin{tabular}[c]{@{}c@{}}Metadata \\ Overhead\end{tabular}} &
  \multirow{2}{*}{\begin{tabular}[c]{@{}c@{}}Performance \\ Overhead\end{tabular}} \\ \cline{3-4}
 &
   &
  \multicolumn{1}{c|}{Inter-Object} &
  Intra-Object &
   &
   &
   &
   \\ \hline
\multirow{8}{*}{\begin{tabular}[c]{@{}c@{}}Tamperproof\\ Metadata\end{tabular}} &
  CHERI &
  \checkmark &
  \checkmark &
  x &
  Prototyped &
  256 bits/pointer &
  $ \propto $ \# of pointer ops \\
 &
  Intel MPX &
  \checkmark &
  \checkmark &
  x &
  \begin{tabular}[c]{@{}c@{}}Discontinued\end{tabular} &
  2 words/pointer &
  $ \propto $ \# of pointer derefs \\
 &
  Hardbound &
  \checkmark &
  \checkmark &
  x &
  Simulation &
  \begin{tabular}[c]{@{}c@{}}0-2 words/pointer,\\ 4 bits/word\end{tabular} &
  $ \propto $ \# of pointer derefs \\
 &
  \begin{tabular}[c]{@{}c@{}}Watchdog\\Watchdoglite\end{tabular} &
  \checkmark &
  \checkmark &
  x &
  Simulation &
  4 words/pointer &
  $ \propto $ \# of pointer derefs/ops \\
 % &
 %  Watchdoglite &
 %  \checkmark &
 %  \checkmark &
 %  x &
 %  Simulated &
 %  4 words per pointer &
 %  $ \propto $ \# of pointer ops \\
 &
  CHEx86 &
  \checkmark &
  x &
  \checkmark &
  Simulation &
  2 words/pointer &
  $ \propto $ \# of pointer ops \\
 &
  REST &
  adjacent only &
  x &
  \checkmark &
  Simulation &
  8-64B token &
  $ \propto $ \# of mem accesses \\
 &
  Califorms &
  \checkmark &
  \checkmark &
  \checkmark &
  Simulation &
  1-7B/pointer &
  $ \propto $ \# of mem accesses \\
 &
  No-FAT &
  \checkmark &
  \checkmark &
  \checkmark &
  Simulation &
  1KB/process table &
  $ \propto $ \# of pointer derefs \\ \hline
\multirow{3}{*}{\begin{tabular}[c]{@{}c@{}}Tamperable\\ Metadata\end{tabular}} &
    \begin{tabular}[c]{@{}c@{}}ARM MTE\\SPARC ADI\end{tabular} &

   % &
  \checkmark &
  x &
  \checkmark &
  \begin{tabular}[c]{@{}c@{}}Commercial\end{tabular} &
  \begin{tabular}[c]{@{}c@{}}Embed inside pointer and \\ 4 bits/16B and 64B objects\end{tabular} &
  $ \propto $ \# of pointer derefs \\
 % &
 %  SPARC ADI &
 %  \checkmark &
 %  x &
 %  \checkmark &
 %  Commercialized &
 %  \begin{tabular}[c]{@{}c@{}}Embed inside pointer and \\ 4 bits per 64B objects\end{tabular} &
 %  $ \propto $ \# of pointer derefs \\
 &
  AOS &
  \checkmark &
  x &
  \checkmark &
  Simulation &
  8B/pointer &
  $ \propto $ \# of pointer derefs \\
 &
  C3 &
  \checkmark &
  x &
  \checkmark &
  Simulation &
  Entangle with pointer &
  $ \propto $ \# of pointer derefs \\ \hline
\end{tabular}

\begin{tabular}{|c|c|cccc|c|c|c|}
\hline
\multirow{2}{*}{Classification} &
  \multirow{2}{*}{Proposal} &
  \multicolumn{4}{c|}{Protected Assets} &
  \multirow{2}{*}{Maturity} &
  \multirow{2}{*}{\begin{tabular}[c]{@{}l@{}}Metadata\\ Overhead\end{tabular}} &
  \multirow{2}{*}{\begin{tabular}[c]{@{}l@{}}Performance \\ Overhead\end{tabular}} \\ \cline{3-6}
 &
   &
  \multicolumn{1}{l|}{RET} &
  \multicolumn{1}{l|}{Code Ptr} &
  \multicolumn{1}{l|}{Data Ptr} &
  Data &
   &
   &
   \\ \hline
\multirow{3}{*}{\begin{tabular}[c]{@{}l@{}}Tamperproof\\ Metadata\end{tabular}} &
  ARM BTI &
  x &
  x &
  x &
  x &
  Commercial &
  1 instruction/branch target &
  $ \propto $ \# of branches \\
 &
  Intel CET &
  \checkmark &
  x &
  x &
  x &
  Commercial &
  \begin{tabular}[c]{@{}l@{}}1 instruction/branch target,\\ 1 word/return address\end{tabular} &
  $ \propto $ \# of branches and returns \\
 &
  ZERO &
  \checkmark &
  \checkmark &
  \checkmark &
  \checkmark &
  Simulation &
  2 bits/word &
  $ \propto $ \# of mem accesses \\ \hline
\multirow{2}{*}{\begin{tabular}[c]{@{}l@{}}Address Layout\\ Randomization\end{tabular}} &
  ASLR &
  \checkmark &
  \checkmark &
  \checkmark &
  \checkmark &
  Commercial &
  none &
  negligible \\
 &
  Morpheus &
  \checkmark &
  \checkmark &
  \checkmark &
  \checkmark &
  Simulation &
  2 bits/word &
  $ \propto $ 1/re-randomization interval \\ \hline
\multirow{2}{*}{\begin{tabular}[c]{@{}l@{}}Tamperable\\ Metadata\end{tabular}} &
  Stack Canary &
  \checkmark &
  x &
  x &
  x &
  Commercial &
  8B/return address &
  $ \propto $ \# of returns \\
 &
  ARM PA &
  \checkmark &
  \checkmark &
  \checkmark &
  x &
  Commercial &
  Embed inside pointer &
  $ \propto $ \# of pointer derefs \\ \hline
\end{tabular}
\caption{\rev{Tradeoffs of Memory Safety Schemes. First table for preactive schemes, second table for reactive schemes. }}
\label{tab:safety}
\end{center}
\end{table*}

In addition to the complicated trade off introduced by varying assumptions about the security of the software, hardware, and attacker's end goals, memory-safety-mechanism designs need to consider the trade-offs in the extent of memory safety provided, the performance overhead, as well as intrusivity of the designs in the hardware and software.

Memory-safety schemes can broadly be categorized into two categories depending on whether they are \emph{preactive} or \emph{reactive} under a memory safety violation. 
Preactive-memory-safety schemes enforce memory safety rules such as spatial safety and/or temporal safety directly. For example, to enforce spatial safety, these mechanisms check whether each memory access is within bounds of the respective (sub-)object. 
To enforce such strict security rules, these mechanisms maintain and operate on a large amount of additional metadata for each pointer, such as base and bound of the pointer for spatial safety, or an allocation identifier for temporal safety.
Thus, preactive-memory-safety schemes tend to have a high performance and storage overhead.

Instead of enforcing memory safety rules directly, reactive-memory-safety mechanisms aim at enforcing relaxed security rules in order to keep the associated overheads low.
Such relaxed security rules allow memory safety violations such as spatial or temporal violations, but aim to protect the confidentiality or integrity for a  selected set of assets in the program. 
For example, stack smashing protection allows a buffer overflow violating spatial safety, but enforces integrity of the return address by later checking the value of the canary.
We summarize the trade-offs among preactive-memory-safety schemes and reactive-memory-safety schemes in \cref{tab:safety}. 
Due to the subjective nature of hardware and software intrusivity, and the difficulty to put in comparison, we instead provide a column on the maturity of the designs. 
}

\vspace{+2ex}
\section{Related Work}
\label{sec:related}

In this work, we conducted a systematic analysis of memory corruption mititigations, distinctively focusing on vulnerabilities that arise at the convergence of memory corruption and side-channel threat models. Prior works have also conducted extensive systematic analysis, albeit focusing on only one of the two traditional threat models.
% \rev{
% We start by first discussing those.}

%We now discuss related work not covered so far.

% \paragraph{Other SoK work}
%As both memory corruption vulnerabilities and micro-architectural side channel attacks are well-studied research problems, there have been high-quality systematic analysis and modeling work targeting each of them separately.
%However, none of these work have looked at the intersections of the two threat models.

\paragraph{SoK on Memory Corruption Attacks and Defenses}
Szekeres et al.~\cite{szekeres2013sok} systematically analyze memory corruption attacks and defenses and
Burow et al.~\cite{burow2017control} systematically compare various proposed CFI (control-flow integrity) mechanisms.
Cowan et al.~\cite{cowan2000buffer} survey the various types of buffer overflow vulnerabilities, attacks, and defenses, and discuss different combinations of prevention techniques.
Saito et al.~\cite{saito2016survey}
survey and classify memory corruption mitigation technologies that are pervasive in operating systems and compilers and
Novković~\cite{novkovic2021taxonomy} analyzes the root causes of memory corruption vulnerabilities and categorizes existing defensive mechanisms based on the attack
techniques they focused on preventing.

\paragraph{SoK on Microarchitectural Side Channels}
DAWG~\cite{dawg} presents a general attack schema of micro-architectural covert and side channels and 
CaSA~\cite{casa} presents a communication-based model to describe side channels.
Pandora's box~\cite{pandora-box} systematically studies side-channel vulnerabilities caused by various micro-architectural optimizations and 
Transient Fail~\cite{transientfail} conducts a systematic analysis of transient execution attacks and defenses.
%How secure is your cache~\cite{he2017secure} propose a probabilistic information flow graph (PIFG) to model the interactions between attackers and victims in cache-based side-channel attacks and He et al.~\cite{he2021new} propose an attack-graph model for reasoning about speculative execution attacks. 
\del{
How secure is your cache propose a probabilistic information flow graph (PIFG) to model the interactions between attackers and victims in cache-based side-channel attacks 
and He et al. propose an attack-graph model for reasoning about speculative execution attacks.} 
%They define a concept, ``security dependency'', between a resource access and its prior authorization operation.
%\mengjia{not sure whether we want to elaborate on the graph model or not.... maybe not, due to space.}
Lastly, Deng et al.~\cite{deng2019analysis} propose a three-step modeling approach to exhaustively enumerate all the possible cache timing-based vulnerabilities.

\rev{
\paragraph{Graph-based Models}
Graph-based models are commonly used to represent side-channel attacks that exploit speculative execution~\cite{inspectre, blade, mosier22, ponce2022, he2017secure, he2021new}.  
At a high level, SIF graphs largely differ from prior works in that they analyze the leakage of the metadata of a defense, while prior works aim to analyze the leakage of program data. 
The closest related work is He et al.~\cite{he2021new}. 
%They aim to develop a graph-based model to represent a variety of speculative execution attacks and are most similar to us in terms of how they define the node types. 
%They develop a graph-based model to represent %a variety of 
%speculative execution attacks and define the node types in a similar way than us. 
There are two key differences. \cite{he2021new} focuses on information leakage from a protected node (often a load). 
In comparison, SIF graphs focus on the information leakage from a security check node (and hence its possible abuse by an attacker).
This key difference ultimately leads to two contradicting outcomes. \cite{he2021new} shows that leakage happens due to the \textit{absence} of security dependencies. To the contrary, SIF graphs show leakage happens due to the \textit{existence} of security dependencies (\cref{sec:model:seq}). 
%They develop a graph-based model but focus on information leakage from a protected node (often a load) while we look a information leakage from a security check node. As a result, \cite{he2021new} shows that leakage happens due to the \textit{abscence} of security dependencies while we show leakage happens due to the \textit{existence} of security dependencies.

}

\rev{
\paragraph{Speculative control-flow hijacking}
%It is important to distinguish our synergistic attacks from speculative control-flow hijacking, such as SpectreRSB~\cite{speculative-bufferoverflow, SpectreRSB} and SpecROP~\cite{specrop}.
It is important to distinguish \ssb attacks from speculative control-flow hijacking~\cite{specrop, speculative-bufferoverflow} attacks.
Speculative control-flow hijacking attacks also combine memory corruption attack techniques to \emph{speculatively} execute attacker-chosen code paths, to leak information via microarchitectural side-channels.
% The control-flow can be manipulated via poisoning branch history buffer~\cite{spectre}, branch target buffers (BTB), return address buffer (RSB)~\cite{koruyeh2018spectre}, and store-to-load forwarding~\cite{speculative-bufferoverflow}.
For example, SpecROP~\cite{specrop} showed that by poisoning BTB, the attacker can speculatively chain small code gadgets to construct more powerful information leakage attacks.
Moreover, Mambretti et al.~\cite{specrop} demonstrated that speculative control-flow hijacking can speculatively bypass stack smashing protection~\cite{stackguard_canaries}, GCC's vtable verification~\cite{vtv}, and Go's runtime memory safety checks~\cite{go-memorysafety}.

There are two key differences between \ssb attacks and speculative control-flow hijacking attacks.
First, speculative control-flow hijacking attacks, by nature, are still ultimately speculative.
The speculative gadgets will eventually be squashed.
Conversely, \ssb attacks use the leaked secrets to bypass memory corruption defenses non-speculatively, ultimately conducting memory corruption attacks that result in architectural consequences.
Second, speculative control-flow hijacking attacks leak the program data. To the contrary, \ssb attacks leak the metadata necessary for memory corruption attacks.
}

\section{Conclusion}
\label{sec:conclusion}

%Synergistic attacks exploit vulnerabilities that span across multiple layers of the system stack. Existing strategies for memory corruption mitigation tend to deal with the non-trivial complexity in today's modern systems by narrowly considering only memory corruption vulnerabilities.
This paper performed a systematic analysis of state-of-the-art memory corruption defense proposals from both industry and academia against \ssb attacks. %that exploit vulnerabilities that arise due to synergies in memory corruption and side-channel vulnerabilities. 
By systematizing a taxonomy we identified the key source of vulnerability to \ssb attacks, namely \emph{spoofable security checks}. The taxonomy also helped us identify two classes of memory corruption mitigations that perform spoofable security checks, i.e., those that use tamperable metadata or randomize the address layout.
Next, we developed a graph-based model that helps us precisely visualize the information flow between the security checks and observable microarchitectural events. The model helps us identify patterns leading to side-channel leakage as well as reason about countermeasures. 
In short, our key contribution is a systematic analysis of memory corruption defenses, focusing on \ssb attacks.

%and categorizing the 20 defenses using this taxonomy, we found 10 vulnerable schemes, among which 5 are either already deployed, or soon-to-be deployed in products by the industry.
%Next, we identified two attack vectors for exploiting the vulnerabilities found in the taxonomy, dissected the defense proposals vulnerabilities individually, and demonstrated three proof-of-concept attacks.
%Our analysis will be useful in guiding researchers to design future memory safety mechanisms that are resistant to synergistic attacks.
%We believe that  Our findings have important implications for designers striving to implement future memory-safe processors and systems.

\newpage

\bibliographystyle{plain}
\bibliography{references}

\end{document}